\documentclass[12pt,preprint]{aastex}

\newcommand{\FtoB}{\mbox{$L_{\mathrm{FIR}}/L_{\mathrm{B}}$}}
\newcommand{\LFIR}{\mbox{$L_{\mathrm{FIR}}$}}
\newcommand{\LAFE}{\mbox{$L_{\mathrm{AFEs}}$}}
\newcommand{\um}{\mbox{$\,\mu{\rm m}$}}
\newcommand{\eg}{e.g.,~}
\newcommand{\ie}{i.e.,~}
\newcommand{\etal}{et al.~}
\newcommand{\kms}{\mbox{$\,$km~s$^{-1}$}}
\newcommand{\ISO}{\footnote{Based on observations with {\it ISO}, an ESA project with 
	    instruments funded by ESA member states (especially the PI countries:
	    France, Germany, the Netherlands, and the United Kingdom) and with 
	    the participation of ISAS and NASA.}}

\begin{document}

\title{Infrared Emission of Normal Galaxies from 2.5 to 12 Microns: {\it ISO} Spectra, 
       Near-Infrared Continuum and Mid-Infrared Emission Features\ISO}

\author{Nanyao Lu\altaffilmark{2}, 
       George Helou\altaffilmark{2}, 
       Michael W. Werner\altaffilmark{3},
       Harriet L. Dinerstein\altaffilmark{4}, 
       Daniel A. Dale\altaffilmark{2, 5}, 
       Nancy A. Silbermann\altaffilmark{2}, 
       Sangeeta Malhotra\altaffilmark{2, 6}, 
       Charles A. Beichman\altaffilmark{3}, 
       Thomas H. Jarrett\altaffilmark{2}
        }

\altaffiltext{2}{Infrared Processing and Analysis Center, MS 100-22, 
	         California Institute of Technology, Pasadena, CA 91125; 
		 lu@ipac.caltech.edu, gxh@ipac.caltech.edu,
		 nancys@ipac.caltech.edu, jarrett@ipac.caltech.edu}
\altaffiltext{3}{Jet Propulsion Laboratory, MS 233-303, 4800 Oak Grove Road,
		 Pasadena, CA 91109; mwerner@sirtfweb.jpl.nasa.gov, chas@mail1.jpl.nasa.gov}
\altaffiltext{4}{Astronomy Department, The University of Texas at Austin, 1 University Station C1400, 
                 Austin, TX 78712-0259; harriet@astro.as.utexas.edu}
\altaffiltext{5}{Dept. of Physics and Astronomy, University of Wyoming, Laramie, WY 82071; ddale@uwyo.edu}
\altaffiltext{6}{Dept. of Physics and Astronomy, Johns Hopkins Univ., Baltimore, MD 21218; san@tarkus.pha.jhu.edu}

\begin{abstract}
\indent
We present ISO-PHOT spectra of the regions 2.5--4.9\um\ and 5.8--11.6\um\ 
for a sample of 45 disk galaxies from the U.S. {\it ISO} Key Project 
on Normal Galaxies. The galaxies were selected to span the range in 
global properties of normal, star-forming disk galaxies in the local 
universe.  The spectra can be decomposed into three spectral components: 
(1) continuum emission from stellar photospheres, which dominates 
the near-infrared (2.5--4.9\um; NIR) spectral region; (2) a weak NIR
excess continuum, which has a color temperature of $\sim 10^3\,$K, 
carries a luminosity of a few percent 
of the total far-infrared dust luminosity \LFIR, and most likely 
arises from the ISM;  and (3) the well-known broad emission features at 
6.2, 7.7, 8.6 and 11.3\um, which are generally attributed to aromatic 
carbon particles.  These aromatic features in emission (AFEs) dominate 
the mid-infrared (5.8--11.6\um; MIR) part of the spectrum, and resemble 
the so-called Type-A spectra observed in many non-stellar sources and 
the diffuse ISM in our own Galaxy.   The few notable exceptions include 
NGC$\,$4418, where a dust continuum replaces the AFEs in MIR, and 
NGC$\,$1569, where the AFEs are weak and the strongest emission feature
is [\ion{S}{4}] 10.51\um.

The relative strengths of the AFEs vary by 15--25\% among the galaxies.  
However, little correlation is seen between these variations and 
either {\it IRAS} 60\um-to-100\um\ flux density ratio 
$R(60/100)$ or the far-infrared-to-blue luminosity ratio \FtoB,
two widely used indicators of the current star-formation activity,
suggesting that the observed variations are not a consequence of 
the radiation field differences among the galaxies.
We demonstrate that the NIR excess continuum and AFE emission are 
correlated, suggesting that they are produced by similar mechanisms 
and similar (or the same) material.  On the other hand, 
as the current star-formation activity increases, the overall strengths 
of the AFEs and the NIR excess continuum drop significantly with 
respect to that of the far-infrared emission from large dust grains. 
In particular, the summed luminosity of the AFEs falls from 
$\sim 0.2\,\LFIR$\ for the most ``IR-quiescent'' galaxies to 
$\sim 0.1\,\LFIR$\ for the most ``IR-active'' galaxies.  This is likely 
a consequence of the preferential destruction in intense radiation 
fields of the small carriers responsible for the NIR/AFE emission.
\end{abstract}


\section{Introduction} \label{sec1}

The spectroscopic properties of galaxies in the wavelength range 
2.5--12\um\ are much less well known than in the optical regime, 
where stellar emission dominates, and in the far-infrared, 
where thermal continuum emission from cool interstellar dust
dominates. 
Prior to the {\it Infrared Space Observatory} ({\it ISO}; Kessler 
\etal 1996) mission,  spectroscopic information on this wavelength 
region was available for only a few of the brightest galaxies and galaxy 
nuclei, which were observed with ground-based telescopes through 
a few spectral windows (\eg Roche \etal 1991), or with the {\it IRAS} 
Low-Resolution Spectrometer (\eg Cohen \& Volk 1989).
However, this relatively unexplored spectral region contains 
important signatures of interstellar dust particles,
in particular the broad emission features at 3.3, 6.2, 7.7, 8.6,
and 11.3\um\ (Gillett \etal 1973) which in the past were often 
called the ``Unidentified Infrared'' or UIR features. These features   
are now known to be ubiquitous in the ISM of our Galaxy: they are seen
in \ion{H}{2} and PDR regions (\eg C\'esarsky \etal 1996; Roelfsema 
\etal 1996; Verstraete \etal 1996, 2001); planetary nebulae and
circumstellar regions (\eg Beintema \etal 1996); reflection nebulae 
(\eg Boulanger \etal 1996, Uchida, Sellgren, \& Werner 1998); 
and diffuse cirrus clouds (\eg Mattila \etal 1996; Lemke \etal 1998;
Chan \etal 2001).  Taken together, they radiate a 
significant fraction of the total IR emission from these regions
(\eg Puget \& L\'eger 1989).

While the exact identity of the carriers of the UIR features is still
unresolved, it is generally agreed that the features arise from 
vibrational modes of a carbon-based, aromatic material, so we will 
refer to them hereafter as aromatic features in emission, or ``AFEs.'' 
Possible candidate materials range from Polycyclic Aromatic Hydrocarbon
molecules (hereafter PAHs; L\'eger \& Puget 1984; Allamandola, 
Tielens, \& Barker 1985, 1989; Puget \& L\'eger 1989) to Hydrogenated
Amorphous Carbon grains (Duley \& Williams 1981, 1988).  
The current picture is that the particles responsible for the emission, 
whether present as free molecules or being attached to larger grains, 
are transiently heated by the absorption of single UV photons to $T 
\sim 10^3\,$K.  The carriers of the AFEs may also play an 
important role in regulating the physical conditions in the ISM by 
contributing significantly to photoelectric heating of the gas (Bakes 
\& Tielens 1994; Helou \etal 2001; however, for an alternate point
of view, see Chan \etal 2001).

The spectral region also hosts a possible near-infrared, non-stellar
continuum emission with a color temperature of $\sim 10^3\,$K, which
was first detected in Galactic reflection nebulae (Andriesse 1978; 
Sellgren, Werner, \& Dinerstein 1983; Sellgren 1984).  Recent observations 
with {\it COBE} and {\it IRTS} hint that this near-IR continuum may also be present 
on larger scales in our Galaxy (Bernard \etal 1994; Tanaka \etal 1996),
but it has not been previously reported in external galaxies.

With its unprecedented sensitivity and contiguous wavelength coverage, 
{\it ISO} made it possible to obtain 2.5--11.6\um\ spectra of large 
numbers of galaxies that were too faint to observe previously.  As part
of an {\it ISO} Key Project to study the physical
properties of the ISM in galaxies (Helou \etal 1997; Dale \etal 2000), 
we obtained 2.5--11.6\um\ spectra using the PHT-S 
mode of ISO-PHOT (Lemke \etal 1996) for 45 galaxies.
This survey covered the full range of morphological 
types of disk galaxies, S0 to Im, that are 
powered by star formation.
Similar {\it ISO} spectra of other    
types of galaxies can be found elsewhere: \eg AGNs (Clavel \etal 2000), 
ultraluminous IR galaxies (Genzel \etal 1998; Rigopoulou \etal 1999),
and galaxies of moderately low surface brightnesses such as 
the Magellanic Clouds (Reach \etal 2000; Sturm \etal 2000; Vermeij 
\etal 2002).  The AFEs are seen in the spectra of 
most of these objects, but are absent or too weak to detect 
in elliptical galaxies (Lu \& Hur 2000; 2003; Athey \etal 2002) and 
extremely metal deficient dwarfs (\eg Thuan, Sauvage, \& Madden 1999).

In Paper I (Helou \etal 2000), we presented spectra for a subsample of
7 galaxies, highlighted the detections of the AFEs and a near-infrared,
non-stellar continuum emission, and suggested that the averaged spectrum
of a number of galaxies could provide a useful template for redshift 
determinations of distant star-forming galaxies.  In the present 
paper, we explore more fully the diversity of the 2.5--12\um\ spectra
of the galaxies in our sample. We present and compare the individual
spectra of all the 45 observed galaxies, provide a quantitative analysis 
of various spectral components, evaluate variations from galaxy to 
galaxy and possible statistical correlations with galaxy properties 
at other wavelengths, and discuss possible reasons for these trends.  
Throughout this paper, we use NIR and MIR to refer to the 2.5--4.9\um\ 
and 5.8--11.6\um\ spectral regions respectively, and AFE(3.3), AFE(6.2),
AFE(7.7), AFE(8.6) and AFE(11.3) for the corresponding individual 
features.

\section{Galaxy Sample, Observations and Results}\label{sec2}

\subsection{The Sample} \label{sec2.1}

The galaxies studied in this paper are a subset of a larger sample 
observed for the Key Project.  This parent sample consists 
of 69 ``normal'' galaxies selected to capture the great diversity in
the properties of galaxies in the local universe, especially in terms
of the ratio of current to past star-formation rate. For each galaxy 
in this project we obtained at least one of the following: 
ISO-CAM images at 7 and 15\um\ (see Dale \etal 2000),
a sparsely sampled ISO-LWS spectrum between 43 and 200\um\ 
targeting fine-structure lines (Malhotra \etal 2001 and references 
therein), and a PHT-S 2.5--11.6\um\ spectrum as presented here.

Table~1 lists the 45 Key Project galaxies for which we obtained a PHT-S 
spectrum. The position of each observation in the table was 
reconstructed from the {\it ISO}
pointing history (IIPH) file, which in most cases was the same as the
intended position.  Nearly all of the observations were taken at
the galaxy optical center, except for two off-center positions in 
NGC~1569 which correspond to emission peaks on the CAM images 
(Hunter \etal 2001).
From Table~1 it can be seen that the blue luminosity $L_B$ ranges
from 2 $\times 10^9$ 
to $7 \times 10^{10}\,L_{\odot}$, the FIR-to-blue luminosity ratio 
\FtoB\ from 0.2 to 15, and the {\it IRAS} 60-to-100\um\ color 
index $R(60/100)$ 
from 0.3 to 1.3.  The heliocentric velocities [Col.~(10)]
were used to shift all spectra to a common (rest) frame for easier 
comparison.  The median velocity of the sample is about 1800\kms.

Fig.~1 shows the distribution of the galaxies in a plot of $\log \FtoB$
{\it vs.} $R(60/100)$. Most of the galaxies fall along a rough diagonal 
from the lower left to upper right corner of the diagram, a trend which
we attribute to an increasing ratio of present-day star forming activity
to the time-averaged star-formation rate in the past (\eg Helou 1986). 
In this paper we assume
that both axes in Fig.~1 are statistically valid measures of the global
amount of current star-formation activity in a galaxy, and thus of 
the UV-to-optical spectral shape of the radiation field.
We define three galaxy subsamples: an ``FIR-quiescent'' subsample, represented
by open squares in Fig.~1; an ``FIR-active'' group, represented by filled squares;
and an intermediate subsample, shown as crosses.  The few 
outliers in Fig.~1 include the compact galaxy NGC$\,$4418, which has
higher $R(60/100)$ than 99\% of the galaxies in the {\it IRAS} Bright Galaxy 
Sample (Soifer \etal 1989), and the low-metallicity irregular galaxy 
NGC$\,$1569 which has low \FtoB.  Most of the outliers also show 
peculiarities in their PHT-S spectra; we comment on these individually 
in \S3.2.

\subsection{Observations and Data Reduction}\label{sec2.2}

The PHT-S spectrometer has two 64-element linear Si:Ga detector arrays. 
The SS array covers the wavelength range  2.5--4.9\um\ with
spectral resolution element 0.04\um\, while the SL array covers the
interval 5.8--11.6\um\ with a resolution element of 0.1\um.
The instrument has a 
$24''\times 24''$ aperture on the sky, pointed with an accuracy $\leq 2''$ 
(Kessler \etal 1996).  For each galaxy, the SS 
and SL spectra were obtained simultaneously, through an aperture 
placed at the position given in Table~1.
Sky reference observations were taken 
at symmetrically placed offsets of $\pm~150''$ from the galaxy center 
along a direction determined by the spacecraft roll angle 
at the time of the observation.
Integration times were 512 seconds, 
split evenly between the galaxy and sky positions, with a duration of
64 sec per chopper step, except for the faint object
IC$\,$860, for which we increased the
total integration time to 2048 seconds.

The spectra were derived from the Edited Raw Data using 
standard procedures in the PHOT 
Interactive Analysis package (PIA version 7; Gabriel \etal 1997), 
including deglitching at both ramp and signal levels, ramp slope 
fitting, signal averaging per chopper position, and sky subtraction of 
the average signal for the two sky reference positions.  The flux calibration
was performed using a signal-dependent ``detector response function'' 
obtained from chopped observations of calibration stars 
with known SED's.  This ``direct calibration,'' 
which was later incorporated into pipeline version 8.4, 
included an empirical correction for the signal loss due to a detector
transient induced by the chopper switching between the source and 
the reference positions.

Since we used point-source flux standards, our spectra
represent the effective emission corresponding to the integration  
of a normalized PHT-S beam profile over the surface brightness
distribution of the source, divided by $f_{psf}$, 
the fraction of the point-source spread function within 
the PHT-S aperture (see Appendix A).   For the SS detector pixels, 
the median flux uncertainty $\sigma^{median}_{SS}$ $\sim 25\,$mJy  
and depends only weakly on the source flux.  In contrast,   
$\sigma^{median}_{SL}$ varies from 15 mJy to 45 mJy
over the sample and scales roughly linearly with source brightness. 
According to the data validation report from pipeline version 8.4,
the absolute and relative flux uncertainties are on order of 
10\% for point sources.  As described in Appendix A, we performed
an independent check on the flux calibration by comparing our PHT-S
data with the CAM imaging data of Dale \etal (2000); these agree
to better than 18\% for 5--8.5\um\ and $\le$ 25\% for the 4--5\um\ 
region.

\subsection{Near-Infrared Photometry}\label{sec2.3}

We used near-infrared images from the Large Galaxy Atlas of the 2-Micron
All Sky Survey, 2MASS (Jarrett \etal 2003) to derive integrated
$J$ (1.25\um), $H$ (1.65\um), and $Ks$ (2.17\um) 
magnitudes for each galaxy, appropriate to the placement 
of the PHT-S aperture on the sky during the {\it ISO} observations. 
Since there is not yet a finalized conversion formula from
the 2MASS magnitude scale to flux densities, we normalized our  
2MASS magnitudes to the multi-aperture observations of  
four early-type galaxies (NGC 4374, NGC 3379, NGC 5866, and NGC 1326)
published by Frogel \etal (1978) and Persson, Frogel, \& Aaronson (1979),
using the magnitude-to-flux conversions 
of Wilson, Schwartz, \& Neugebauer (1972). 
To adjust our 2MASS photometry to the magnitude scale
in these references, we found it necessary to add the following offsets:
(0.028 $\pm$ 0.009)$^{m}$ at $J$,  
(0.052 $\pm$ 0.004)$^{m}$ for $H$, and (0.079 $\pm$ 0.010)$^{m}$ to convert
from $K_s$ to $K$ (2.2\um). 
The resulting 2MASS fluxes are used to normalize our PHT-S spectra in 
\S3 and \S4 to unveil the NIR excess continuum in disk galaxies. 
This requires no significant 
zero-point offset between the 2MASS and PHT-S flux scales.  We show in 
Appendix~B that this is indeed the case.

\subsection{Results}\label{sec2.4}

The sky-subtracted, rest-frame PHT-S spectra are presented in Figures~2$a$-2$e$.
Table~2 summarizes the PHT-S aperture coverage factor $p$ (see Appendix A), 
the mean continuum flux density at 4\um, and a mean flux density for each 
of the AFEs.  These mean flux densities, defined in the footnotes to Table~2, 
are derived after shifting the spectrum to the rest frame and resampling 
the data at the PHT-S detector wavelengths by a linear interpolation between 
the two nearest data points.  Note that the mean flux density for AFE(11.3)
is basically taken over only the blue side of the feature, since part of 
the red side fell outside our wavelength coverage.   The last 2 lines in Table~2 
provide information on two elliptical galaxies, NGC~3379 and NGC~4374, whose 
PHT-S spectra were discussed by Lu \& Hur (2000; 2003), and which served as 
comparison objects for the disk galaxies (\S 4.1).

\section{The Spectra} \label{sec3}

\subsection{General Characteristics} \label{sec3.1}

The great majority of the observed galaxies show qualitatively similar
PHT-S spectra. The MIR part of the spectra is typically dominated by 
the prominent AFE features at 6.2, 7.7, 8.6, and 11.3\um, which have 
relative strengths and profiles similar to those of ``Type~A'' Galactic
sources as described, for example, by Geballe (1997) and Tokunaga (1997).  
These features are superposed on a NIR continuum dominated by the emission
from stellar photospheres.
Of all the spectra, only that of NGC$\,$4418 display MIR structures 
that are qualitatively different from a Type~A spectrum.  We discuss 
in more detail this unique spectrum and a few others in \S3.2.

The similarity of most of the spectra in the range 5.8--11.6\um\ suggests
that it is useful to derive an averaged spectrum to serve as a template 
for normal, star-forming galaxies.  We derived such an average spectrum
from 40 of the 45 galaxies in this study, omitting only the atypical 
galaxies NGC~4418 and NGC~1569 (see above, and \S 3.2), and the three 
galaxies with the lowest S/N ratios in their spectra (NGC$\,$3705, 
NGC$\,$4519, and NGC$\,$7418).
The average spectrum was obtained on a wavelength pixel-by-pixel
basis from the resampled, rest-frame individual spectra (see \S2.4)
normalized at some fiducial wavelength.  
No correction for redshift was made to the {\it JHK} points described in 
\S 2.3, since the effects are negligible for these broad photometric 
bands.

The averaged spectra resulting from two choices of the fiducial 
wavelength for normalization are shown in Fig.~3.  The open squares 
depict the spectrum obtained by normalizing the individual spectra
at {\it J}, while the thick solid curve corresponds to normalizing by 
the integrated flux of AFE(7.7).   In both cases we used 
$1/\sigma^2$-weighted averaging, 
where $\sigma$ is either 
$\sigma^{median}_{SS}$ or $\sigma^{median}_{SL}$ depending on which 
detector array segment the pixel under consideration belongs to;
the {\it JHK} data points were averaged using the same $\sigma$-weights 
as used for the PHT-SS data.

The error bars in Fig.~3 represent $s/\sqrt{n}$, where $s$ is 
the standard deviation of the normalized fluxes and $n$ is 
the number of galaxies included in the average  
(due to the redshifts of individual galaxies, 
$n < 40$ for some points near the array edges).  For a given pixel, 
$s$ is given by
$$ s^2 = (\Sigma\,w_if^2_i - <f>^2\Sigma\,w_i)/({n-1 \over n}\Sigma\,w_i),
	 \eqno(1)$$
where $f_i$ and $w_i$ are respectively the flux density and weight from 
the $i$th sample galaxy, and $<f>$ is the average flux of the pixel.
Therefore, these error bars reflect mainly the variation of spectral 
shape  within the sample.   
Note that the two normalization methods
have different biases: normalization at 7.7\um\ gives 
greater weight to individual spectra which are
more dominated by emission from the ISM (\eg the AFEs), whereas
normalization at {\it J} gives greater weight to galaxies with a larger 
proportional contribution of starlight.

The averaged spectra in Fig.~3 can be compared with the ``template'' spectrum
presented in Fig.~2 of Paper I. The latter was a
straight average [\ie with $w_i \equiv 1$ in eq.~(1)] over a 
subset of 28 galaxies. In addition, a different PHOT calibration
and a different normalization parameter (the strength of the 
6.2\um\ AFE) was used.  The main differences are limited to 
$\lambda < 3$\um; the MIR region of the template
spectrum from Paper I is very similar to those derived here if
they are all normalized in the same way.

We also compare our average disk galaxy spectra with those of the two
elliptical galaxies, NGC$\,$3379 and NGC$\,$4374. 
The PHT-S spectra of these two E1 galaxies, reduced from the {\it ISO}
archive data by Lu \& Hur (2003), are shown in Fig.~3 as a thin solid line 
and dotted line respectively.
At a heliocentric redshift of $v_h = 911$\kms, NGC~3379 (M$\,$105) is one 
of the nearest normal giant elliptical galaxies, with a classical
r$^{\onequarter}$ profile (de Vaucouleurs \& Capaccioli 1979).  
NGC~4374 (M$\,$84), at $v_h = 1060$\kms, is known to host a 
central radio source (cf. Bridle \& Perley 1984), and contains at least some
interstellar matter, indicated by 
the presence of dust lanes and {\it IRAS} detections at 60 and 100\um.
Nevertheless, these two ellipticals have quite similar
PHT-S spectra, nearly featureless and falling  
roughly as a Rayleigh-Jeans law from 1.25 to $\sim$ 7\um. An apparent 
flattening beyond 8\um\ may be due to circumstellar dust emission 
(Knapp, Gunn, \& Wynn-Williams 1992; Athey \etal 2002).

It can be seen from Fig.~3 that the spiral galaxies show clear excess 
emission compared to the ellipticals at all wavelengths $\ge$ 2.2\um. 
In Table~3, we list the numerical values of the two average disk-galaxy
spectra in Fig.~3, as well as the average for the two E galaxies.
These averaged spectra, because of their high S/N, offer a more sensitive
way to study the profiles of the brighter AFEs and to look for weak 
features that might be buried in noise in individual spectra.

Table~4 summarizes the features that are identifiable in the averaged 
spectra in Fig.~3 and were discussed in Paper I. 
The equivalent widths of some of weaker features were estimated on
the AFE(7.7)-normalized spectrum and are given in the last column of
Table~4.  The 3.3\um\ feature, which is not apparent in most of 
the individual galaxy spectra, is easily recognized in the averaged 
spectrum, and the narrow feature at around 4.03\um\ is probably 
\ion{H}{1} Br$\alpha$.  The hump near 7.0\um\  
could be either [\ion{Ar}{2}] 6.99\um\ and/or the 6.9\um\
dust feature 
discussed by Bregman \etal (1983) and Cohen \etal (1986), with possibly 
a small contribution from H$_2$ 6.910\um\ $v$=0--0 S(5).
The small bump at around 10.6\um\ could arise from 
[\ion{S}{4}], which is seen most distinctly in
the post-starburst galaxy NGC$\,$1569.
Finally, the asymmetric appearance of AFE(7.7)
is probably a result of the blending of two unresolved
peaks at 7.6 and
7.8\um; such substructure has been seen in some
Galactic sources
(\eg Bregman 1989; Roelfsema \etal 1996), and is 
partially resolved in some of the highest-S/N spectra in this study
(e.g. NGC$\,$4194 and IC$\,$883, see Fig.~2).

In order to assess whether the the MIR spectral shape varies 
with star-formation activity, we compared the average spectra 
for the FIR-quiescent and FIR-active subsamples defined in 
\S 2.1, which include 14 galaxies each.
The average spectra for the $J$-band normalization described above,
aligned at {\it J} for comparison, are shown in Fig.~4$a$ as the
solid curve (FIR-quiescent) and open squares (FIR-active).
It can be seen from the figure that both the NIR continuum
and AFEs for the FIR-active galaxies lie well above those of 
the FIR-quiescent subsample.
On the other hand, when the averages are normalized by
AFE(7.7) and aligned at 7.7\um\ (Fig.~4$b$), 
the FIR-quiescent spectrum lies above the FIR-active one 
only in the NIR region.  (The small difference between 
the two spectra in Fig.~4$b$ in the region 9--10\um\ is 
significant only at the $1.5\,\sigma$ level.)
Note that the apparent ``dips'' around 3.2, 3.7 and 4.1\um\ seen 
in Fig.~4$b$ are probably not real, but are due to the
AFE(7.7) normalization, which yields an overly strong weighting
for the (noisy, and in some cases negative) short-wavelength 
data points of galaxies with strong AFEs and faint NIR continua.

These plots demonstrate that the only significant spectral 
difference between the two subsamples lies in the NIR, and 
can be attributed to different contributions from starlight. 
In other words, the spectral shape of the MIR AFEs in galaxies 
remains largely independent of star-formation activity.

\subsection{Remarks on Individual Spectra} \label{sec3.2}

\subsubsection{NGC 520}

NGC$\,$520 (Arp 157), classified as an
intermediate-stage merger by Hibbard \& van Gorkom (1996), is as radio- and
infrared-bright as NGC$\,$4038/4029 (Arp 244; the Antennae).  
Numerical simulations suggest that NGC$\,$520 is a merger remnant resulting 
from two disk galaxies which began colliding about
300 million years ago (Stanford \& Balcells 1991).   
Our PHT-S aperture was placed 
on the main optical ridge, $\sim$ $14''$ north of 
the emission peak in the 10\um\ image of Bushouse, Telesco, \& Werner (1998).
This peak is presumably the dust-obscured
nucleus of the brighter galaxy in the pair (see Fig.~6 of Stanford \& 
Balcells 1990). The PHT-S spectrum therefore included both this
obscured nucleus and parts of the disks of both galaxies.

The most striking aspect of the spectrum of NGC$\,$520 is the 
weakness of AFE(8.6) and AFE(11.3) relative to AFE(7.7). 
On the other hand, AFE(6.2) has normal (Type~A)
strength relative to AFE(7.7).
In the context of the PAH model, both AFE(8.6) and AFE(11.3) arise
primarily from C-H bending modes, while AFE(6.2) and AFE(7.7) 
come from C-C modes.  Thus, a plausible conjecture is that the compression
of the ISM due to the interaction
of the merging galaxy disks
might have led to a high degree of dehydrogenation 
of PAHs, resulting in relatively weak 8.6 and 11.3\um\ features.

\subsubsection{NGC 1569}

NGC$\,$1569 is a nearby Magellanic-type irregular galaxy with a metallicity 
$\sim$ 30\% of solar.  It experienced a strong burst of star formation 
as recently as a few
million years ago (Israel 1988; Israel \& van Driel 1990; Greggio \etal 1998). 
We obtained PHT-S spectra at three positions on the galaxy disk.  Position~C 
(see Table~1) is on the nucleus,    
which also hosts one of the two superluminous young star clusters in this galaxy,
Cluster B (Ables 1971; Arp \& Sandage 1985).  
Positions NW and SE are located  ($16\farcs0$ W, $5''$ N) and  
($16\farcs6$ E, $7''$ S) of the nucleus respectively. 
These correspond respectively to the NW and SE peaks on our CAM LW2 image 
(Hunter \etal 2001) as well as to \ion{H}{2} regions No.~2 and No.~7
in Table~3 of Waller (1991).

All three spectra show a strong, unresolved emission line at 10.55 
($\pm 0.05$)\um, which presumably is [\ion{S}{4}] 10.51\um.
We may also have detected [\ion{Ar}{3}] 8.99\um, in the NW and SE 
spectra.  If present, it is a factor of 5--10 weaker than [\ion{S}{4}] 10.51\um, 
indicating a relatively high effective temperature for the illuminating 
radiation field (\eg Rank \etal 1978; Rubin 1985). This is 
consistent with the conclusions of Hunter \etal (2001), who infer 
$T_{eff}$ = 40,000 K based on the Key Project LWS spectroscopy and
other data. The large inferred number of early-type O stars indicates 
recent vigorous star-formation activity in NGC~1569 (Hunter \etal 2001).

The unusual strength of the ionic lines relative to the dust emission in 
NGC~1569 carries over to longer wavelengths.  Its value of
L([\ion{O}{3}] 88\um)/\LFIR, where L([\ion{O}{3}] is the luminosity in 
the [\ion{O}{3}] 88\um\ line emission, 
is higher by an order of magnitude or more (1.0--1.5 dex) than that of 
nearly all of the other galaxies in the Key Project sample (see Fig.~6$a$ 
of Malhotra \etal 2001). The only other galaxy with such high-contrast
ionic line emission is IC~4662, for which, unfortunately, we did not 
obtain a PHT-S spectrum.  Recall also that NGC~1569 was an outlier in Fig.~1, 
having unusually low \FtoB\ for its {\it IRAS} color index $R(60/100)$.
Whether this is due to the recent starburst or is a consequence of 
low metallicity is unclear, but the weak dust compared to the gas emission 
suggests a low dust-to-gas ratio in NGC~1569, and therefore possibly a high 
mean free path for far-UV photons as well as a hard UV radiation field.

\subsubsection{NGC 4418}

The PHT-S spectrum of NGC$\,$4418 is markedly different from those of 
nearly all the other sample galaxies: the AFEs are not seen, and the MIR 
spectrum is dominated by an apparent emission plateau extending from 
6--9\um, which has been interpreted as arising from a broad continuum 
upon which is superposed a deep 10\um\ silicate absorption feature 
(Roche \etal 1986).

NGC~4418 was previously noticed to be an extreme object by Malhotra \etal (1997),
who pointed out its extreme deficiency in the [\ion{C}{2}] 157\um\ line.
It is a good example of the now well-established trend, 
that L([\ion{C}{2}] 157\um)/\LFIR\ 
to \LFIR\ decreases as the FIR luminosity and intensity of the dust-heating radiation 
field increases (Malhotra \etal 1997; Luhman \etal 1998). 
The nature of the illuminating radiation
field in NGC~4418 is unclear, although its  
FIR color index, $R(60/100) = 1.3$, is 
the highest (``warmest'') in the full Key Project sample
(see Table~5 of Malhotra \etal 2001).
NGC$\,$4418 is a very compact source at radio wavelengths, with a radius of 
$\le$ $0\farcs5$ ($\sim 53h^{-1}\,$pc) at 20$\,$cm (Condon \etal 1990), so
our PHT-S spectrum enclosed most (84\%) of its MIR emission.
On the one hand, our PHT-S spectrum of NGC~4418 resembles the {\it ISO} spectrum 
of an ultra-compact \ion{H}{2} region in M17 (C\'esarsky \etal 1996), which 
would be consistent with the suggestion that NGC~4418 contains an intense 
nuclear starburst of very high optical depth.  On the other hand, 
Spoon \etal (2001) report seeing absorption
features from ices in the PHT-S spectrum, and interpret
NGC~4418 as a dust- and ice-enshrouded 
active galactic nucleus (AGN).

\subsubsection{IC 860}

IC$\,$860 is also compact, with a 20$\,$cm diameter of  
$< 0\farcs4$ (or 100$h^{-1}$pc; Condon \etal 1990),
consequently, our PHT-S aperture included about
81\%\ of its MIR emission.   
This galaxy has been generally classified as a non-AGN (\eg Leech 
\etal 1989).  Given its compact size, warm FIR color of 
$R(60/100) = 0.94$, and high value of $\log \FtoB = 1.05$, 
it is plausible
that IC$\,$860 harbors a nuclear starburst.  
While the AFEs in IC$\,$860 have typical relative strengths,
they have a low collective intensity relative to the FIR dust emission
(see \S4.3). IC~860 was also one of the 
extremely ``[\ion{C}{2}]-deficient'' galaxies noted by Malhotra \etal (1997).
If the carriers of the AFEs are major contributors of photoelectrons
that heat the gas, these two observations taken together are consistent
with the destruction of 
a substantial fraction of the AFE carriers
by an intense UV radiation field.

\subsubsection{NGC 5866}

NGC$\,$5866 is a nearby, early-type (S0), edge-on disk galaxy, with a dust
lane visible in optical and the ``coolest'' FIR color
index in the entire Key Project sample, $R(60/100) = 0.3$
The CAM LW2 image (Malhotra \etal 2000) shows that the emission at 
7\um\ arises mainly from the edge-on disk of the galaxy. Our PHT-S 
aperture was centered roughly on the nucleus, and enclosed about
60\% of the total flux of the galaxy.

The PHT-S spectrum of NGC~5866 shows a strong continuum 
rising towards shorter wavelengths, consistent with the picture that 
the global emission from this galaxy is strongly
dominated by stellar photospheric emission.
There appears to be a broad emission feature that peaks 
at 7.9\um, instead of the usual AFE(7.7\um) feature.  
The 8.6\um\ feature is weak, if present, and the 11.3\um\ feature 
is much wider than usual.  Taken together, these AFEs are unusually
weak relative to the FIR dust emission as compared to the other 
FIR-quiescent galaxies in the sample (see \S4.3 below).

\section{Analysis and Discussion} \label{sec4}

\subsection{Continuum Emission at 3 to 5 Microns}\label{sec4.1}

Spectral synthesis models indicate that as long as the luminosity 
of a galaxy is dominated by relatively old stellar populations 
(\ie older than a few Gyrs), 
the shape of its stellar continuum spectrum in near-infrared 
remains largely independent of the details of its star formation
history (\eg Bruzual \& Charlot 1993).
This is supported by results from 2MASS, which show   
that normal galaxies display a much smaller color dispersion 
in the near-infrared than in the optical (\eg Jarrett 2000). 
We therefore assume that the averaged elliptical galaxy spectrum   
(\S 3.1) is representative of the intrinsic spectral
shape of the stellar component in our disk galaxies.
It is uncertain whether the circumstellar dust emission longward 
of 8\um\ can also be scaled in this way, but for disk galaxies, 
the emission is much fainter than the ISM emission in this spectral
region.

We first investigated whether internal reddening by dust can account
for the observed NIR excess emission.  
In Fig.~5, the same 40 galaxies\footnote{In \S4 we use galaxies drawn 
from this 40-galaxy sample only.  In comparing PHT-S fluxes with 
the FIR fluxes (\ie in Tables~5 and 7, and Fig.~9), we further 
exclude those few without a PHT-S aperture coverage factor $p$ in 
Table~2.} included in the average spectrum (see \S3.1) 
are shown in two color-color plots: logarithmic flux-density
ratios $K/J$ {\it vs}. $H/J$, and f(4\um)$/J$ 
{\it vs}. $H/J$, respectively.  
The galaxies are coded according to their degree of FIR-activity
in the same way as in Fig.~1, and the elliptical galaxies are
shown as open circles.
We considered two geometries for the dust extinction: a foreground
dust screen (shown as solid lines in Fig.~5); and the case of 
uniformly mixed stars and dust (shown as dotted curves).
In both cases, we used a near-infrared reddening law of $A(\lambda)/A(J) = 
(\lambda/1.25\mu{\rm m})^{-1.7}$, after Mathis (1990). 
We assumed zero reddening for the ellipticals.  
The tick marks indicate 
A$_V$ = 1, 2, and 3$^{m}$ respectively, for 
the dust screen model; in the star/dust mixed case, one requires 
about 3 times as much total dust column density in order
to produce the same amount of reddening as for the foreground screen.
Apart from this scale factor, the two cases produce 
very similar trends in Fig.~5 within 
the parameter space occupied by our galaxies.
As one can see from the figure,
most of the galaxies lie above the reddening lines, particularly
in the lower panel, f(4\um)$/J$ {\it vs}. $H/J$. Therefore, it
appears that dust reddening cannot account  
for the color differences 
between our sample of disk galaxies and 
the reference elliptical galaxies. There must
be an additional continuum component
which becomes more prominent at the longer wavelengths.

We set out to reconstruct the character of this ``NIR excess continuum'' 
as follows. We assumed that the {\it J} and {\it H} emission in the disk galaxies 
arises only from stars. Next, we subtracted the averaged elliptical
galaxy spectrum in Table~3 from the data for each disk galaxy,
after reddening it to the $H/J$ color of the latter using the foreground
screen case.
(For the few cases where the observed $H/J$ color is slightly bluer than 
that of the ellipticals, we take A$_V$ = 0.)
Finally, to increase the S/N ratio, we averaged
the residual spectra for the entire galaxy sample. The result is shown in 
Fig.~6$a$. Also shown is a modified black-body 
curve for $T = 750\,$K and a $\lambda^{-2}$ emissivity law, 
which fits the general shape of the curve quite well.
The emission is clearly much broader and stronger than AFE(3.3\um),
and therefore is unlikely to be simply the wings of this feature.

The residual spectrum shown in Fig.~6$a$ represents a lower limit to 
the typical intensity of this NIR excess continuum, because of
our assumption that it does not contribute at H.
In order to set an upper limit to such a contribution,
we considered the effects of assuming that only the $J$-band
flux has no contribution from non-stellar sources, and derived the
residual spectrum that results from subtracting the
elliptical galaxy spectrum uncorrected for different
reddening between the disk and E galaxies.
This yields the average residual spectrum shown in Fig.~6$b$, 
which is well fit by a modified black-body curve with $T = 10^3\,$K 
and a $\lambda^{-2}$ emissivity, only slightly hotter
than the curve in Fig.~6$a$. 
Both approaches therefore indicate that the color temperature 
of the non-stellar NIR continuum is close to $10^3\,$K.
The fractional energy contained in the NIR spectral region is about 
10\% in Fig.~6$a$ and about 17\% in Fig.~6$b$.

Next we consider whether the NIR excess continuum is correlated
with AFE emission. In Fig.~7, we plot the ratio of the
average 4\um\ emission (taken to be representative of the
NIR excess emission) to the mean flux density of AFE(7.7),
against $\log \FtoB$.
(The two methods of removing the stellar continuum, discussed
above, lead to more or less the same 4\um\ fluxes.)  
The strength of AFE(7.7\um) relative to
the $J$-band flux increases by nearly an order of magnitude
within the sample (see Table~2), as \FtoB\ increases.
On the other hand, there is no net trend in Fig.~7, indicating
that the NIR excess correlates strongly with the MIR emission
for galaxies with a wide range in dust content and luminosity.
We view this as strong evidence that the NIR excess continuum 
originates in the ISM, as opposed to being (scattered) starlight
from late-type stars or circumstellar dust emission.   However, 
the scatter in Fig.~7 is greater than can be accounted for by 
measurement errors, suggesting that there are some real variations
among galaxies in the ratio of the NIR excess to AFE(7.7). 
The most deviant point is the early-type galaxy NGC~5866,
which, as mentioned in \S 3.2.5, has very weak AFE features.

This non-stellar dust continuum is probably the same component 
that has been detected in reflection nebulae and the large-scale
ISM of our own Galaxy (see \S1).  In fact, Sellgren \etal (1985) 
showed that, for the reflection nebulae NGC$\,$7023 and NGC$\,$2023, 
the 4\um\ emission surface brightness is between 1/6 to 1/10 of 
that at the peak of AFE(7.7).  This range is quantitatively 
consistent with our results in Fig.~6. It has been suggested that such a NIR 
continuum might be due to electronic fluorescence or 
a quasi-continuum of overlapping bands of PAH molecules (\eg Allamandola, 
Tielens, \& Barker 1989).  While further work is needed in order
to determine whether the AFEs and NIR continuum emission
arise from exactly the same macromolecules or material,
it seems clear that the carriers of these two components are closely 
related.

In Table~5 we summarize the luminosity in the NIR region relative to 
the FIR dust emission, for each of the three subsamples defined in \S2.1.  
The total NIR luminosity is $5-18\%\,\LFIR$ and arises mostly from   
stellar photospheres.  If we correct for the stellar contribution 
using the two methods described above, 
the non-stellar NIR continuum contributes 
$\sim 3-4\%\,\LFIR$ for the most quiescent
galaxies in our sample, and $1-2\%\,\LFIR$ for the most FIR-active ones.
These estimates are on the same order as the {\it COBE} results for our own Galaxy
(Bernard \etal 1994).

\subsection{Variations in the Mid-Infrared AFEs} \label{sec4.2}

Some variations have been observed among Galactic sources
in the profiles of the 
AFEs (\eg Roelfsema \etal 1996; Peeters \etal 2002) and
the ratios of one feature relative to another
(\eg Joblin \etal 1996; Lu 1998; Vermeij \etal
2002), but the physical
implications are still not fully understood.
In particular, it appears that relative strengths
of the features are quite insensitive to the intensity
and color temperature of the 
local radiation field (\eg Uchida \etal 2000; 
Chan \etal 2001). 
Likewise, while we find similar variations in the
AFEs in our sample of galaxies, 
we have not found any statistically significant
correlation between these variations and  
\FtoB, $R(60/100)$, optical morphology, 
the 7\um-to-15\um\ flux density ratio (Dale \etal 2000), 
or the mean surface brightness at 7 or 15\um\ within an
isophotal ellipse containing 50\% of the infrared flux.

In Fig.~8 we plot the ratios of the strengths of the 6.2, 8.6, 
and 11.3\um\ AFEs to AFE(7.7) against \FtoB\ for the 40 galaxies,
which all show clearly discerned AFEs.
We hereafter abbreviate the mean $F_{\lambda}$
of a feature by its wavelength, e.g. (6.2) for the mean 
flux of AFE(6.2).  There is some indication of a slight 
decreasing trend in these plots.
However, these trends are weak and probably not statistically significant. 
We computed the median feature strength ratios separately for 
the  FIR-quiescent, intermediate, and FIR-active 
subsamples defined in \S 2.1. The values are given in Table~6 
and plotted as large crosses in Fig.~8. 
It can be seen that the median values decline by less
than 1 $\sigma_s$ from FIR-quiescent to FIR-active,
where $\sigma_s$ is the r.m.s.~dispersion
of the entire sample (see the last row of Table~6). 
This differs from the conclusion of Lu \etal (1999) that
the ratio (11.3)/(7.7) decreases with increasing \FtoB.
That earlier study was based on an older PHT-S calibration,
but more importantly, the stellar continuum was removed
in a more simplistic way, by subtracting a power-law fit 
to the spectral intervals 3.5--5\um\ and 9.5--10.5\um.
That procedure had the effect of undersubtracting the continuum 
underlying the 11.3\um\ feature in galaxies with a strong stellar
component, thus producing an apparent trend.

The scatter in Fig.~8 is significantly greater than the 
statistical error bars, implying that there are intrinsic 
galaxy-to-galaxy variations in the relative feature strengths. 
According to Table~6, the spread 
in the AFE ratios is $\sim 15\%$ for (6.2)/(7.7) and (8.6)/(7.7),
and $\sim 25\%$ for (11.3)/(7.7).  
Within the framework of the PAH hypothesis, variations in 
the relative feature strengths have been attributed 
to the presence of different fractions of PAHs that are 
ionized and/or hydrogenated, as a consequence of differences
in the ambient UV radiation field (\eg Jourdain de Muizon 
\etal 1990; de Frees \etal 1993; Schutte \etal 1993; 
Langhoff 1996). However, our data suggest that the situation 
may be more complex than a simple dependence on the UV 
radiation field.  It is plausible that chemical processing may 
also play an important role in determining the local abundance 
and emission properties of the PAHs (\eg Boulanger \etal 1990).

The spectra in Fig.~2 also show some galaxy-to-galaxy variations 
in the AFE profiles.  In Appendix C, we quantify one aspect of
these variations by defining a ``logarithmic slope'' $S$ that 
measures the steepness of the feature profile on the short- or 
long-wavelength side.  As we show in Table~C1, the galaxy-to-galaxy 
variations in $S$ range from 15--30\% for all the AFEs, except 
for the short-wavelength side of AFE(8.6) which shows possibly 
much greater variations relative to the median value.  
This could either be due to intrinsic 
variations in the profile of AFE(8.6), or to a contribution from 
a weak feature around 8.2\um\ that varies from galaxy to galaxy.  
Such a shorter-wavelength
feature was invoked by Verstraete \etal (2001) to fit high-S/N 
ISO-SWS spectra of Galactic sources, and an emission feature
at $\sim$ 8.2\um\ has also been observed in some post-AGB stars 
(Peeters \etal 2002).

\subsection{The Destruction of the AFE Carriers} \label{sec4.3}

It was previously known that the AFEs are depressed in \ion{H}{2}
regions (\eg C\'esarsky \etal 1996).  This has been widely 
attributed to the preferential destruction of the small AFE 
carriers relative to larger dust grains in these regions.  {\it 
IRAS} data also indicate an AFE depression over galaxy scales, 
with the AFEs being more severely depressed in galaxies with 
warmer $R(60/100)$ colors (Helou, Ryter, \& Soifer 1991).

Fig.~9 shows how $L_{AFEs}$, the summed luminosity of the four MIR AFEs
(6.2, 7.7, 8.6, and 11.3\um), correlates with \LFIR.
The few galaxies without a PHT-S aperture coverage factor $p$ in
Table~2 are not plotted here.  We indicate galaxies with values 
of $p < 35\%$ (for which the aperture correction is large and 
therefore less certain) by filled symbols, in order to distinguish 
them from the others (open symbols).   For some of these galaxies, 
Malhotra \etal (2001, Table~C1) have derived $G_0$, the intensity 
of the far-UV interstellar radiation field in units of the radiation 
field in the solar neighborhood, based on a PDR model that depends 
on the observed fluxes in [\ion{C}{2}] 158\um\ and [\ion{O}{1}] 
63\um and the flux of the total infrared emission.
These galaxies are plotted as squares of three different sizes, 
where the smallest corresponds to $2.3 \le \log G_0 < 
3.4$, the intermediate size to $3.5 \le \log G_0 < 3.9$, and 
the largest to $4.0 \le \log G_0 < 4.8$.

For comparison, we also show in Fig.~9 the {\it IRAS} results of Werner, 
Gautier, \& Cawlfield (1994) for multiple spatial positions in 
two Galactic \ion{H}{2} regions, the Rosette and California nebulae  
(indicated by a dashed line), and the reflection nebula in the Pleiades 
(indicated by a solid line).   These lines actually represent 
the logarithmic ratio of the {\it IRAS} 12\um\ flux (assumed to be 
proportional to our integrated AFE flux) to the FIR flux, and are 
shifted vertically in order to bracket most of the galaxy data points.
The average trend of our galaxy points is steeper than 
that characteristic of the reflection nebula, but shallower than the trend
for the \ion{H}{2} regions.  This comparison suggests that both \ion{H}{2}
regions and reflection nebulae contribute to the total AFE and FIR 
emission of galaxies.

In Table~7 we list median ratios of several different measures of
the AFE overall intensity to the FIR flux, for the three subsamples. 
It can be seen from the table that the depression of AFEs relative 
to the FIR emission becomes significant only for the most FIR-active 
subsample; for the latter galaxies, \LAFE/\LFIR\ has a value
only 60\%\ that of the other two subsamples.  Furthermore, from 
Fig.~9 it can be seen that this AFE depression is more related to 
the hardness than the intensity of the radiation field: at a given 
$R(60/100)$, galaxies with higher values of $G_0$ tend to lie closer
to the dashed line.  Since the flux ratio of [\ion{C}{2}] to the AFEs
has very little dispersion in this sample of galaxies (Helou \etal
2001), one does expect such a linkage between the diagnostics based
on the fine-structure lines and the AFEs.  However, this is unlikely
just a mathematical linkage, for the two reference lines in 
Fig.~9 clearly suggest that the (UV-hard) HII regions have a lower 
\LAFE/\LFIR\ than the (UV-soft) reflection nebulae at any given 
$R(60/100)$.

If the photodestruction process that weakens the AFEs in Galactic 
\ion{H}{2} regions is also responsible for depressing the AFE 
emission on galaxy scales, our results imply that the mass spectrum
of the dust particles could differ significantly between quiescent
and starburst galaxies.

As shown in \S 3.1, the MIR emission from disk galaxies is
dominated by strong AFE emission over a wide range in such physical
characteristics as the intensity of the interstellar UV radiation field,
FIR luminosity, and temperature of the FIR-emitting dust. The few 
exceptions we find may be indicative of the extreme conditions required
for the AFEs to fade from view. Representing the high-radiation-field 
limit, we have objects such as NGC~4418 (which may be an AGN), IC~860,
and NGC~1569 (which, if plotted in Fig.~9 despite its small value for
$p$, would lie near the position of IC~860). For these objects, it
is reasonable to explain the weakness of the AFEs as being the
consequence of destruction of the AFE carriers by hard UV photons
(or shocks), by analogy with the Galactic sources discussed above.

It is less clear what causes the deficiency in AFE emission in
the FIR-quiescent galaxy NGC~5866. This is the earliest type 
galaxy in Key Project sample for which we have a PHT-S spectrum;
unfortunately, we did not obtain PHT-S observations for 
the other three E/S0 galaxies in our program 
(Malhotra et al. 2000). The lack of strong AFE emission
in NGC~5866 could be due either to an actual deficiency of the
AFE carriers or to a lack
of energetic UV photons capable of exciting the 
features, or both. Note that our LWS observations
of the [\ion{C}{2}] 158\um\ and [\ion{O}{1}] 63 \um\
lines indicate a particularly weak and soft UV
radiation field in this galaxy (Malhotra et al. 2000).

\section{Summary} \label{sec5}

We present new {\it ISO} PHT-S spectra (2.5--11.6\um), extended to 
1.25\um\ using {\it JHK} data from the 2MASS survey, for a sample
of 45 disk galaxies that span the typical range in global properties
of galaxies energetically dominated by star formation.  PHT-S aperture
coverage factors are also provided for most of the galaxies.
We decompose the spectra into three constituents:
(1) stellar continuum emission, which dominates at the
shortest wavelengths (2.5--4.9\um; NIR); 
(2) a weaker and redder NIR ``excess continuum'';
and (3) the well-known aromatic dust emission features 
(AFEs) at 6.2, 7.7, 8.6, and 11.3\um\, which dominate 
the long-wavelength half of the PHT-S spectra (5.8--11.6; MIR).
Most of the galaxy spectra appear similar to each other,
especially in the MIR region, and to an averaged ``template'' 
spectrum shown in Fig.~3.  The most striking exceptions are 
NGC$\,$4418, which shows no AFEs but instead has a broad
mid-infrared continuum upon which is superposed a 
strong 10\um\ silicate absorption feature; and NGC$\,$1569,
which has a very weak infrared continuum but an unusually 
strong [\ion{S}{4}] 10.5\um\ line from ionized gas.

The non-stellar NIR excess continuum has an average color 
temperature of $\sim 10^3\,$K and a luminosity of a few 
percent of \LFIR, where \LFIR\ is the far-infrared 
(40--120\um) luminosity attributed to large dust grains.  
This NIR continuum scales more or less linearly with 
the strength of the AFEs, suggesting that the NIR excess 
continuum originates in the ISM of galaxies (and not, 
for example, in late-type stars or circumstellar dust);
and that the AFEs and NIR excess continuum arise from similar 
carriers.

The profiles and relative strengths of the AFEs in the disk 
galaxies match those of ``Type A'' mid-infrared spectra, 
by far the most predominant type of mid-infrared dust 
emission pattern in the ISM of our own Galaxy.  This resemblance 
suggests that the carriers of the Type-A AFEs are also prevalent
in the ISM of other galaxies.  The combined luminosity of 
the AFEs in the region 5.8--11.3\um\ is 10--20\%$\,$\LFIR.
The relative strengths of the AFEs vary on average by 
15--25\%.  These observed variations, however, do not 
correlate with {\it IRAS} 60\um-to-100\um\ flux density 
ratio, $R(60/100)$, or the far-infrared-to-blue luminosity 
ratio \FtoB, two commonly used indicators of 
global star-formation activity in galaxies. We interpret this
as indicating that other factors, in addition to the present-day
radiation field, affect the strengths and shapes of the AFEs.

The ratios of both the AFEs and non-stellar NIR continuum 
to the FIR  flux decrease systematically from the most quiescent 
galaxies to the most actively star-forming galaxies in 
the sample.  We show that this is more related to the hardness
than the intensity of the heating radiation field, likely
a result of the AFE/NIR carriers being preferentially destroyed,
relative to the larger dust grains responsible for the FIR 
emission, in regions of active star formation and intense UV 
radiation fields. This implies that the mass and size distribution 
functions of the interstellar dust particles vary in environments 
with different levels of star formation activity.

\acknowledgments

We are grateful to J. A. Acosta-Pulido for his help on 
the PHT-S flux calibrations, and to the other Key Project members, 
J. Brauher, A. Contursi, D. Hollenbach, D. A. Hunter,
M. Kaufman, S. Kolhatkar, 
K.-Y. Lo, S. D. Lord, R. H. Rubin, G. J. Stacey, and H. A. Thronson, 
Jr., ~for their contributions to the project from which this paper 
was derived.  We thank the anonymous referee for a number of comments 
that improved the presentation of this paper.
The {\it ISO} Photometer, ISO-PHOT, was built by Lemke \etal (1996).  
The data presented here were reduced with the ISOPHOT Interactive Analysis
(PIA) package, a joint development by the ESA Astrophysics Division 
and the ISOPHOT Consortium with the collaboration of the Infrared 
Processing and Analysis Center (IPAC). This work was supported in part 
by {\it ISO} data analysis funding from the National Aeronautics and Space 
Administration (NASA), and has made use of the NASA/IPAC Extragalactic
Database (NED) which is maintained by the Jet Propulsion Laboratory, California
Institute of Technology, under contract with NASA.

\appendix 

\section{Flux Comparison of the PHT-S and ISO-CAM Data} \label{appendixA}

ISO-CAM LW2 images are available for 42 galaxies in Table~1 from Dale 
\etal (2000), and for one additional galaxy, NGC$\,$5866, in the ISO
archive from the observations described by Vigroux \etal (1999).
For each of these galaxies, we integrated the LW2 filter curve over the PHT-S
spectrum to derive $f_p(LW2)$, a CAM LW2-equivalent flux density at 
6.7\um.  The LW2 bandpass extends to about 5\um\ on the short-wavelength side.  
For its wavelength coverage beyond the blue end of the SL array, we 
used a simple linear interpolation between the red end of the SS array
and the blue end of the SL. 

We then centered the PHT-S aperture on the CAM image at the position 
of our PHT-S observation.  Because a CAM filter-wheel jitter can 
affect the effective pointing of CAM images,
a small residual positional uncertainty remains.
After subtracting a constant sky level from the CAM image,
we derived the quantity $f_c(LW2)$ by integrating the source surface brightness
over the PHT-S aperture.  As a result, we have
$$f_c(LW2) = \int_{24''\times 24''} SB(x,y)dxdy, \eqno(A1)$$
where $SB(x,y)$ is the source surface brightness distribution function.  
We can express $f_p(LW2)$ in a similar way:
$$f_p(LW2) = (1/f_{psf})\,\int B(x,y)\,SB(x,y)dxdy, \eqno(A2)$$
where $f_{psf}$ is the fractional point-source spread function for PHT-S and $B(x,y)$ 
is the PHT-S beam profile (which has a value of unity near the aperture center).
We have implicitly assumed that $f_{psf}$, $B(x,y)$, and $SB(x,y)$
are all effective values over the PHT-S detector pixels included 
within the CAM LW2 bandpass.  The integration in eq.~(A1) is over the PHT-S 
aperture only, while that in eq.~(A2) is over the entire source.  For a 
source that is sufficiently compact that it can be treated as a point 
source with respect to the PHT-S beam profile, one has $f_c(LW2)/f_p(LW2)
\approx f_{psf}$.  For the PHT-S detector pixels relevant to the CAM 
LW2 filter, $f_{psf} \approx 0.92$.  Therefore, for point sources we 
should have $f_c(LW2)/f_p(LW2) \approx 0.92$, if there were no systematic
flux offset between the two instruments. 

Fig.~10 is a plot of $f_c(LW2)/f_p(LW2)$ as a function of $p$, the fractional
CAM LW2 flux of a galaxy that falls within the PHT-S aperture, which is
given by the ratio of $f_c(LW2)$ to the total LW2 flux of the galaxy.
Notice that the scatter is larger for more extended sources (\ie 
smaller values of $p$); this effect may be partly due to the residual positional 
uncertainty described above.  In addition, $f_c(LW2)/f_p(LW2)$ increases on 
average as $p$ drops, presumably due to the beam size effect in eq.~(A1).  
The average flux ratio for the 20 galaxies with $p > 0.6$ is 
(0.76 $\pm 0.03$) and is indicated by the dotted line in Fig.~10.  This implies
a possible systematic flux difference of no greater than 18\% 
between the CAM LW2 and PHT-S results.

A similar comparison was performed with the CAM LW1 (4--5\um) data
(Dale \etal 2000) on the 10 galaxies for which these data were available. 
The result is shown in Fig.~11.  The scatter in this diagram is 
significantly greater than in Fig.~10, largely because the galaxies 
are much fainter in LW1.   Nevertheless, a quantitatively 
similar result is found as for the LW2 comparison:
the median flux ratio in Fig.~11 is 0.7, which implies a difference of about
25\% in flux scale after taking $f_{psf}$ into account.

\section{Zero-Point Difference between the PHT-S and 2MASS Fluxes} \label{appendixB}

Using the 2MASS K-band flux density and the average PHT-S 4\um\ flux density 
as defined in Table~2, we show here that the 
zero-point offset between the 2MASS and PHT-S flux scales is insignificant.
In Fig.~12$a$ we plot $f_{\nu}(4\um)$ against $f_{\nu}(K)$ for our 
sample galaxies.  The dotted line is a least-squares fit to all the data
points, by minimizing their distances perpendicular to the fit, and  
intercepts the vertical axis at about (0.006 $\pm~0.005)$ Jy.
However, we believe that this overestimates the true zero-point offset.
In fact, $f_{\nu}(4\um)/f_{\nu}(K)$ should rise 
somewhat as $\log \FtoB$ increases, because the
contribution of the non-stellar 3--5\um\ continuum emission (see \S 4.1) 
has a greater influence on $f_{\nu}(4\um)$ than on $f_{\nu}(K)$. 
Indeed, Fig.~12$b$ shows 
that there are more FIR-active galaxies at smaller K fluxes.  A better 
estimate of the zero-point offset in flux 
is therefore given by the solid line 
in Fig.~12$a$, which is a similar least-squares fit but includes only
those galaxies with $\log \FtoB < 0.1$ (which are
shown as solid squares in 
both Figs.~12$a$ and 12$b$).  This line has a vertical interception of only 
(0.002 $\pm~0.007)$ Jy.  We therefore conclude that
the zero-point offset between the PHT-S and SMASS flux scale is insignificant.

\section{Slopes of Feature Profiles} \label{appendixC}

For each AFE, we 
define a ``logarithmic slope'' parameter $S$ that measures the steepness
of the feature profile on the short or long- wavelength side,
$$ S = {I_0 - I_1 \over 
	  I_0 + I_1}, \eqno(C1) $$
where $I_0$ is the flux density of the detector pixel nearest 
the peak of the emission feature in the rest-frame, and $I_1$ is 
the flux density of a detector pixel down one side of the feature.
For $I_0$ we used the pixel corresponding to 6.216\um\ in Table~3
for AFE(6.2), 7.616\um\ for AFE(7.7), 8.540\um\ for AFE(8.6) and 
11.263\um\ for AFE(11.3).  For simplicity we chose $I_1$ at one 
of the detector pixels that was used to define the wavelength 
range of the feature (see Table~2).  Two slopes can be defined
for each AFE, $S_{-}$ for the short-wavelength side and $S_{+}$ 
for the long-wavelength side, except for AFE(11.3), for which 
only $S_{-}(11.3)$ can be defined.

In Table~C1, we list for each AFE: the median value for the 40-galaxy 
sample defined in \S 3.1 [Col.~(2)]; the r.m.s.~dispersion around 
this median [Col.~(3)]; and $\sigma_m$, the median of the statistical
measurement errors [Col.~(4)]. 
While the ratio of Cols.~(3) to (2) tells us about the variation
in the slope, the ratio of Cols.~(3) to (4) can be used to gauge 
whether this is significant. The ratio of Cols.~(3) to (2) ranges 
from $\sim$ 15--30\% for both slopes of AFE(6.2) and
AFE(7.7) and for $S_{+}(8.6)$, to 40\% for $S_{-}(11.3)$, to 130\% 
for $S_{-}(8.6)$.  The ratios of Cols.~(3) to (4) show that, of 
these, only the variations in $S_{-}(8.6)$ are significant at 
3$\,\sigma_m$.  Possible reasons for the observed variations in 
$S_{-}(8.6)$ are discussed \S4.2.



%
\begin{deluxetable}{lrrrclrrcr}
\tabletypesize{\scriptsize}
\tablecaption{Galaxy Properties}
\tablenum{1}
\tablewidth{0pt}
\tablehead{
\colhead{Galaxy}   & \colhead{R.A.$^a$}   &  \colhead{Dec$^a$}  & 
\colhead{ROLL$^b$}     & \colhead{ISO-TDT$^c$}  & \colhead{Morphology$^d$} &
\colhead{$\log\,L_B^e$}    & \colhead{$\log\,\FtoB^f$} & \colhead{$R(60/100)^g$} & \colhead{$v_h^h$} \\
\colhead{(1)}   & \colhead{(2)}  & \colhead{(3)}   & \colhead{(4)}  &
\colhead{(5)}   & \colhead{(6)}  & \colhead{(7)}   & \colhead{(8)}  &
\colhead{(9)}   & \colhead{(10)}}

\startdata
   NGC 278            &  05204.6 &  473301 &  73.03 & 59702263  & SAB(rs)b        &  9.77 &  0.01\phn\phn\phn  & 0.54 &  641 \\
   NGC 520 \phn       &  12435.0 &   34742 & 247.28 & 77702280  & Irr             & 10.33 &  0.36\phn\phn\phn  & 0.66 & 2281 \\
   NGC 693            &  15030.9 &   60841 &  67.88 & 59502319  & I0:sp           &  9.61 &  0.13\phn\phn\phn  & 0.60 & 1564 \\
   NGC 695            &  15114.2 &  223456 &  66.50 & 63300751  & IB?(s)m:pec     & 10.85 &  0.50\phn\phn\phn  & 0.58 & 9735 \\
   NGC 1022           &  23832.5 &  -64038 & 244.41 & 78401024  & (R')SB(s)a      &  9.83 &  0.26\phn\phn\phn  & 0.73 & 1503 \\
   UGC 02238          &  24617.4 &  130544 &  72.90 & 63301036  & Pec             & 10.21 &  0.82\phn\phn\phn  & 0.53 & 6436 \\
   NGC 1222           &  30856.8 &  -25717 & 258.14 & 82400843  & S0- pec:        &  9.87 &  0.44\phn\phn\phn  & 0.84 & 2455 \\
   NGC 1317           &  32244.7 & -370609 & 201.48 & 75001077  & SAB(r)a         & 10.03 & -0.40\phn\phn\phn  & 0.34 & 1941 \\
   NGC 1326           &  32356.4 & -362749 & 201.39 & 75001158  & (R)SB(rl)0/a    &  9.84 & -0.28\phn\phn\phn  & 0.58 & 1362 \\
   NGC 1385           &  33728.2 & -243004 & 241.72 & 79600846  & SB(s)cd         & 10.06 & -0.03\phn\phn\phn  & 0.46 & 1493 \\
\\
   UGC 02855          &  34822.9 &  700759 &  94.09 & 62902698  & SB(s)cd II-III  & 10.12 &  0.31\phn\phn\phn  & 0.46 & 1203 \\
   NGC 1482           &  35439.4 & -203007 & 240.50 & 79600986  & SA0+ pec sp     &  9.54 &  0.93\phn\phn\phn  & 0.72 & 1916 \\
   NGC 1546           &  41436.6 & -560338 & 124.41 & 68900662  & SA?a pec        &  9.72 & -0.22\phn\phn\phn  & 0.32 & 1278 \\
   NGC 1569 C         &  43049.1 &  645052 &  90.07 & 64600492  & IBm             &  9.33 & -0.66\phn\phn\phn  & 0.98 & -104 \\
   -------------- NW \phn &  43046.6 &  645057 & 244.83 & 64600492  & ......      &  .... &  ....\phn\phn\phn  & .... & .... \\
   -------------- SE  &  43051.7 &  645045 & 244.87 & 64600492  & ......          &  .... &  ....\phn\phn\phn  & .... & .... \\
   NGC 2388           &  72853.5 &  334905 &  94.06 & 71802365  & SA(s)b: pec     &  9.80 &  1.11\phn\phn\phn  & 0.67 & 4134 \\
   ESO 317-G023       & 102442.4 & -391822 & 328.02 & 25200171  & (R')SB(rs)a     &  9.90 &  0.53\phn\phn\phn  & 0.57 & 2892 \\
   NGC 3583           & 111410.7 &  481901 & 292.74 & 19500259  & SB(s)b          & 10.27 & -0.18\phn\phn\phn  & 0.38 & 2136 \\
   NGC 3620           & 111604.8 & -761252 & 340.02 & 27600983  & (R')SB(s)ab     &  .... &  ....\phn\phn\phn  & 0.70 & 1779 \\
   NGC 3683           & 112732.1 &  565242 & 294.46 & 19401040  & SB(s)c?         &  9.75 &  0.42\phn\phn\phn  & 0.47 & 1656 \\
   NGC 3705           & 113006.8 &   91638 & 295.43 & 18400677  & SAB(r)ab        &  9.68 & -0.62\phn\phn\phn  & 0.33 & 1017 \\
\\                                                                                                 
   NGC 3885           & 114646.6 & -275523 & 307.94 & 25200727  & SA(s)0/a        &  9.86 &  0.04\phn\phn\phn  & 0.71 & 1802 \\
   NGC 3949           & 115341.5 &  475131 & 299.79 & 19500332  & SA(s)bc:        &  9.65 & -0.19\phn\phn\phn  & 0.42 &  798 \\
   NGC 4027           & 115930.6 & -191547 & 299.65 & 24200368  & SB(s)dm         & 10.01 & -0.08\phn\phn\phn  & 0.41 & 1671 \\
   NGC 4102           & 120623.3 &  524240 & 301.90 & 19500586  & SAB(s)b?        &  9.52 &  0.54\phn\phn\phn  & 0.68 &  837 \\
   NGC 4194           & 121410.0 &  543142 & 303.81 & 19401376  & IBm pec         &  9.97 &  0.63\phn\phn\phn  & 0.93 & 2506 \\
   NGC 4418           & 122654.7 &  -05242 & 292.74 & 24100408  & (R')SAB(s)a     &  9.47 &  1.15\phn\phn\phn  & 1.37 & 2179 \\
   NGC 4490           & 123036.8 &  413823 & 299.88 & 20501580  & SB(s)d pec      &  9.91 & -0.21\phn\phn\phn  & 0.59 &  578 \\
   NGC 4519           & 123330.5 &   83916 & 290.87 & 23600331  & SB(rs)d         &  9.46 & -0.30\phn\phn\phn  & 0.53 & 1221 \\
   NGC 4691           & 124813.4 &  -31958 & 292.86 & 23101069  & (R)SB(s)0/a pec &  9.62 & -0.02\phn\phn\phn  & 0.64 & 1110 \\
   IC 3908            & 125640.5 &  -73342 & 293.15 & 25202254  & SB(s)d?         &  9.33 &  0.29\phn\phn\phn  & 0.48 & 1303 \\
\\                                                                                                 
   IC 860             & 131503.5 &  243707 & 281.96 & 61800104  & SB(s)a:         &  9.66 &  1.05\phn\phn\phn  & 0.96 & 3865 \\
   IC 883             & 132035.3 &  340824 & 301.38 & 21501377  & Pec             & 10.23 &  1.14\phn\phn\phn  & 0.69 & 6892 \\
   NGC 5433           & 140236.0 &  323037 & 314.85 & 57100315  & SAB(s)c:        & 10.25 &  0.33\phn\phn\phn  & 0.57 & 4352 \\
   NGC 5713           & 144011.3 &  -01724 & 281.31 & 28400959  & SAB(rs)bc pec   & 10.15 &  0.20\phn\phn\phn  & 0.57 & 1883 \\
   NGC 5786           & 145856.9 & -420045 & 298.09 & 29900767  & (R')SAB(s)bc    & 10.69 & -0.49\phn\phn\phn  & 0.35 & 3054 \\
   NGC 5866           & 150629.4 &  554546 & 273.89 & 26902854  & S0              &  9.79 & -0.60\phn\phn\phn  & 0.30 &  672 \\
   NGC 5962           & 153631.7 &  163632 & 279.95 & 27800783  & SA(r)c          & 10.20 & -0.09\phn\phn\phn  & 0.40 & 1958 \\
   IC 4595            & 162044.2 & -700835 & 268.25 & 27601375  & SB?c sp II:     & 10.28 &  0.14\phn\phn\phn  & 0.39 & 3410 \\
   NGC 6286           & 165831.8 &  585612 & 355.19 & 20700516  & SB(s)0+ pec?    & 10.16 &  0.83\phn\phn\phn  & 0.37 & 5595 \\
   NGC 6753           & 191123.7 & -570256 & 247.95 & 29901232  & (R)SA(r)b       & 10.55 & -0.02\phn\phn\phn  & 0.34 & 3142 \\
\\                                                                                                  
   NGC 7218           & 221011.7 & -163936 & 249.17 & 36902415  & SB(r)c          &  9.88 & -0.17\phn\phn\phn  & 0.42 & 1662 \\
   NGC 7418           & 225635.9 & -370145 & 243.31 & 36902723  & SAB(rs)cd       &  9.97 & -0.32\phn\phn\phn  & 0.33 & 1446 \\
   IC 5325            & 232843.1 & -411957 & 237.82 & 36902824  & SAB(rs)bc       &  9.88 & -0.26\phn\phn\phn  & 0.35 & 1503 \\
   NGC 7771           & 235124.9 &  200641 & 66.54  & 21900879  & SB(s)a          & 10.69 &  0.41\phn\phn\phn  & 0.49 & 4287 \\
   Mrk 331            & 235126.2 &  203508 & 73.52  & 56500644  & SA(s)a: pec     & 10.09 &  1.12\phn\phn\phn  & 0.76 & 5541 \\
\enddata											 
\tablenotetext{a}{J2000 Right Ascension and Declination in ``hhmmss.s'' and ``ddmmss,'' respectively.}
\tablenotetext{b}{The ISO roll angle in degrees, measured in a counterclockwise manner from the celestial north to the spacecraft 
		  Z-axis which is perpendicular to the direction in which the chopper was operated for sky reference positions.
	          This roll angle gives the sky orientation of the square aperture of PHT-S which is aligned with the spacecraft
		  axes.}
\tablenotetext{c}{The TDT number of the ISO observation as it appears in the ISO archive.}
\tablenotetext{d}{Optical morphology taken from Dale \etal (2000) if available, otherwise from the RC3 catalog (de Vaucouleurs
		  \etal 1991).}
\tablenotetext{e}{Logarithmic optical blue luminosity in solar units, derived as in Dale \etal (2000).}
\tablenotetext{f}{Logarithmic FIR-to-blue luminosity ratio, derived as in Dale \etal (2000).}
\tablenotetext{g}{{\it IRAS} 60\um-to-100\um\ flux density ratio, derived from {\it IRAS} addscan fluxes [see Table~4 of Dale \etal (2000)].}
\tablenotetext{h}{Heliocentric velocity in units of \kms.}
\end{deluxetable}

%
\begin{deluxetable}{lcrrrccccc}
\tabletypesize{\scriptsize}
\tablenum{2}
\tablewidth{0pt}
\tablecaption{Mean Flux Densities $F_{\lambda}$ around Selected Wavelengths$^a$}
\tablehead{
\colhead{Galaxy}   & \colhead{$p^b$}   &  \colhead{J$^c$}  & 
\colhead{H$^c$}     & \colhead{K$^c$}  & \colhead{4$\mu$m$^d$}  &
\colhead{(6.2)$^e$}     &  \colhead{(7.7)$^e$}  & \colhead{(8.6)$^e$}  &
\colhead{(11.3)$^e$} \\
\colhead{(1)}   & \colhead{(2)}  & \colhead{(3)}   & 
\colhead{(4)}   & \colhead{(5)}  & \colhead{(6)}   & 
\colhead{(7)}   & \colhead{(8)}  & \colhead{(9)}   & 
\colhead{(10)}}

\startdata
   NGC 278             &  0.31 & 27.46 & 18.48 &  8.27  & 0.87(0.13) &  1.93(0.03) &  2.79(0.02) &  1.39(0.02) &  1.46(0.03)  \\
   NGC 520 \phn        &  0.23 &  9.77 &  8.05 &  4.40  & 0.57(0.13) &  1.33(0.03) &  1.82(0.03) &  0.51(0.03) &  0.21(0.05)  \\
   NGC 693             &  0.72 & 12.63 &  9.45 &  4.70  & 0.92(0.09) &  1.02(0.03) &  1.27(0.02) &  0.73(0.03) &  0.66(0.05)  \\
   NGC 695             &  0.73 &  7.75 &  5.40 &  2.75  & 0.92(0.12) &  1.83(0.03) &  2.72(0.02) &  1.30(0.03) &  \,....\phn(\,....\phn)  \\
   NGC 1022            &  0.78 & 13.09 & 10.23 &  4.95  & 0.65(0.08) &  2.18(0.03) &  3.37(0.02) &  1.68(0.02) &  1.42(0.05)  \\
   UGC 02238           &  0.80 &  5.07 &  4.26 &  2.45  & 0.66(0.11) &  1.91(0.03) &  2.91(0.02) &  1.37(0.04) &  1.15(0.05)  \\
   NGC 1222            &  0.76 &  8.56 &  6.22 &  2.85  & 0.84(0.10) &  1.65(0.03) &  2.29(0.02) &  1.14(0.02) &  1.14(0.03)  \\
   NGC 1317            &  .... & 37.42 & 25.44 & 11.40  & 1.42(0.09) &  0.62(0.02) &  0.73(0.02) &  0.46(0.02) &  0.48(0.06)  \\
   NGC 1326            &  0.70 & 50.08 & 34.37 & 15.54  & 1.30(0.26) &  1.23(0.02) &  1.59(0.01) &  0.89(0.02) &  0.94(0.03)  \\
   NGC 1385            &  0.29 & 13.14 &  8.86 &  3.93  & 0.45(0.07) &  1.13(0.02) &  1.69(0.02) &  0.89(0.02) &  0.80(0.04)  \\
\\                                                                                                                
   UGC 02855           &  0.08 & 16.19 & 13.80 &  7.02  & 0.80(0.13) &  1.04(0.03) &  1.72(0.03) &  0.81(0.03) &  0.80(0.04)  \\
   NGC 1482            &  0.66 & 22.08 & 19.36 & 10.48  & 2.47(0.07) &  5.88(0.03) &  9.33(0.03) &  4.05(0.02) &  3.79(0.04)  \\
   NGC 1546            &  0.33 & 21.24 & 15.66 &  7.21  & 0.84(0.18) &  1.53(0.05) &  2.09(0.04) &  1.09(0.03) &  1.29(0.10)  \\
   NGC 1569 C          &  0.22 & 25.33 & 16.51 &  7.47  & 1.05(0.09) &  0.46(0.02) &  0.35(0.02) &  0.24(0.03) &  0.30(0.06)  \\
   -------------- NW \phn & 0.32 & 18.02 & 11.75 &  5.43 & 0.98(0.10) & 0.82(0.03) &  0.79(0.03) &  0.66(0.02) &  0.96(0.03)  \\
   -------------- SE   &  0.15 & 12.61 &  8.28 &  3.72  & 0.38(0.09) &  0.35(0.03) &  0.38(0.02) &  0.24(0.02) &  0.26(0.04)  \\
   NGC 2388            &  0.84 & 11.19 &  8.75 &  4.40  & 0.69(0.11) &  1.75(0.04) &  2.59(0.06) &  1.24(0.03) &  0.87(0.04)  \\
   ESO 317-G023        &  0.74 & 10.48 &  8.07 &  4.08  & 0.76(0.10) &  1.11(0.03) &  1.88(0.02) &  0.89(0.03) &  0.92(0.06)  \\
   NGC 3583            &  0.37 & 19.57 & 13.82 &  6.40  & 1.20(0.11) &  0.87(0.03) &  1.17(0.02) &  0.60(0.02) &  0.69(0.05)  \\
   NGC 3620            &  0.75 & 30.99 & 27.11 & 14.64  & 2.33(0.11) &  3.95(0.04) &  6.38(0.03) &  2.89(0.02) &  1.78(0.05)  \\
   NGC 3683            &  0.47 & 18.21 & 14.35 &  7.07  & 1.25(0.11) &  2.09(0.04) &  2.90(0.02) &  1.33(0.02) &  0.91(0.06)  \\
   NGC 3705            &  0.20 & 20.28 & 14.60 &  6.57  & 0.85(0.09) &  0.50(0.05) &  0.39(0.03) &  0.20(0.02) &  0.07(0.05)  \\
\\                                                                                                                
   NGC 3885            &  0.82 & 26.73 & 19.72 &  9.51  & 1.83(0.14) &  1.87(0.03) &  2.57(0.03) &  1.30(0.03) &  1.40(0.04)  \\
   NGC 3949            &  0.24 & 14.47 & 10.04 &  4.35  & 0.47(0.10) &  0.77(0.04) &  1.06(0.02) &  0.51(0.02) &  0.55(0.05)  \\
   NGC 4027            &  .... &  7.83 &  5.32 &  2.30  & 0.37(0.08) &  0.69(0.03) &  0.89(0.03) &  0.39(0.02) &  0.31(0.04)  \\
   NGC 4102            &  0.62 & 45.15 & 34.18 & 17.04  & 3.39(0.14) &  4.44(0.03) &  6.78(0.03) &  3.36(0.03) &  2.94(0.06)  \\
   NGC 4194            &  0.81 & 11.84 &  8.62 &  4.27  & 0.42(0.13) &  2.47(0.03) &  3.68(0.03) &  1.68(0.02) &  1.02(0.08)  \\
   NGC 4418            &  0.84 &  6.68 &  4.57 &  2.09  & 0.47(0.11) &  1.35(0.04) &  2.86(0.03) &  1.23(0.02) &  0.29(0.06)  \\
   NGC 4490            &  0.05 & 13.87 &  8.66 &  3.62  & 0.58(0.12) &  0.40(0.03) &  0.41(0.03) &  0.22(0.02) &  0.33(0.03)  \\
   NGC 4519            &  0.27 &  4.62 &  3.17 &  1.31  & 0.39(0.11) &  0.36(0.03) &  0.46(0.03) &  0.25(0.03) &  0.11(0.04)  \\
   NGC 4691            &  .... & 13.09 &  8.70 &  3.94  & 0.91(0.16) &  1.94(0.07) &  2.69(0.04) &  1.39(0.04) &  1.38(0.06)  \\
   IC 3908             &  0.52 & 11.51 &  8.80 &  4.42  & 0.84(0.11) &  1.51(0.03) &  2.14(0.01) &  0.77(0.12) &  0.95(0.05)  \\
\\                                                                                                                
   IC 860              &  0.89 &  5.30 &  3.76 &  1.73  & 0.28(0.03) &  0.20(0.01) &  0.31(0.01) &  0.12(0.01) &  0.16(0.02)  \\
   IC 883              &  0.90 &  3.88 &  2.83 &  1.53  & 0.56(0.12) &  1.26(0.03) &  1.90(0.03) &  0.76(0.02) &  0.43(0.04)  \\
   NGC 5433            &  0.71 &  7.62 &  5.90 &  2.99  & 0.72(0.12) &  1.23(0.05) &  1.90(0.02) &  0.87(0.04) &  0.74(0.07)  \\
   NGC 5713            &  0.45 & 16.58 & 11.34 &  5.26  & 1.66(0.13) &  2.30(0.04) &  3.34(0.03) &  1.64(0.04) &  1.73(0.07)  \\
   NGC 5786            &  0.30 & 11.42 &  8.37 &  3.80  & 0.83(0.11) &  0.66(0.03) &  0.84(0.02) &  0.46(0.03) &  0.39(0.05)  \\
   NGC 5866            &  0.61 & 64.78 & 46.92 & 22.32  & 3.32(0.11) &  0.87(0.04) &  0.68(0.02) &  0.38(0.02) &  0.41(0.04)  \\
   NGC 5962            &  0.30 & 22.20 & 14.70 &  6.71  & 0.53(0.13) &  0.72(0.04) &  1.05(0.02) &  0.61(0.02) &  0.40(0.07)  \\
   IC 4595             &  0.34 &  9.90 &  8.58 &  4.32  & 0.75(0.11) &  1.04(0.04) &  1.45(0.02) &  0.68(0.02) &  0.70(0.05)  \\
   NGC 6286            &  0.69 &  7.71 &  6.41 &  3.55  & 0.90(0.14) &  1.56(0.06) &  2.44(0.02) &  1.00(0.02) &  0.87(0.06)  \\
   NGC 6753            &  0.33 & 38.53 & 28.44 & 13.21  & 1.64(0.10) &  1.19(0.02) &  1.40(0.02) &  0.70(0.02) &  0.82(0.03)  \\
\\                                                                                                                
   NGC 7218            &  0.36 &  8.23 &  5.58 &  2.45  & 0.41(0.10) &  0.58(0.03) &  0.78(0.02) &  0.35(0.02) &  0.35(0.04)  \\
   NGC 7418            &  0.12 &  5.45 &  3.70 &  1.69  &-0.11(0.14) &  0.30(0.03) &  0.32(0.02) &  0.13(0.02) &  0.09(0.04)  \\
   IC 5325             &  0.22 &  9.98 &  6.84 &  3.02  & 0.23(0.08) &  0.48(0.02) &  0.59(0.02) &  0.24(0.02) &  0.23(0.05)  \\
   NGC 7771            &  0.52 & 17.35 & 13.41 &  6.70  & 1.16(0.10) &  1.80(0.04) &  2.74(0.03) &  1.34(0.03) &  1.13(0.07)  \\
   Mrk 331             &  0.76 &  8.98 &  6.72 &  3.46  & 0.80(0.12) &  1.82(0.04) &  2.73(0.02) &  1.32(0.03) &  1.22(0.07)  \\
   NGC 3379            &  .... &119.31 & 79.66 & 34.94  & 4.81(0.06) &  0.86(0.02) &  0.50(0.01) &  0.48(0.01) &  0.39(0.03)  \\
   NGC 4374            &  .... & 99.70 & 71.52 & 31.67  & 3.77(0.06) &  0.61(0.01) &  0.41(0.01) &  0.40(0.01) &  0.36(0.02)  \\
\enddata	
\tablenotetext{a}{All flux densities are in units of $10^{-14}\,{\rm W}\,{\rm m}^{-2}\,\mu{\rm m}^{-1}$.
		  To convert between this unit and Jy, we use 1.25, 1.65, 2.2, 4\um\ as the effective 
	          wavelengths for Columns~(3), (4), (5) and (6), respectively.  The parenthesized value 
		  following a flux density is the statistical error of that flux.}
\tablenotetext{b}{PHT-S aperture coverage factor $p$ as defined in Appendix A.}
\tablenotetext{c}{JHK flux densities within the PHT-S aperture.}
\tablenotetext{d}{The mean 4\um\ flux density derived from a simple average of the flux densities 
		  in Jy of the 27 pixel channels between 3.4 and 4.4\um.}
\tablenotetext{e}{The mean flux density of an AFE, defined as the integrated flux between $\lambda_1$
		  and $\lambda_2$ divided by $(\lambda_2 - \lambda_1)$, where $\lambda_1$ and $\lambda_2$
	          are respectively 5.98\um\ and 6.64\um\ for AFE(6.2), 7.20\um\ and 8.22\um for 
		  AFE(7.7), 8.22\um\ and 9.23\um for AFE(8.6), and 10.86\um\ and 11.40\um for 
		  AFE(11.3).}
\end{deluxetable}

%
\begin{deluxetable}{rrrrr}
\tabletypesize{\scriptsize}
\tablenum{3}
\tablewidth{0pt}
\tablecaption{Rest-Frame Average Spectra$^a$}
\tablehead{
			      \colhead{$\lambda$($\mu$m)}
			      & \colhead{Norm.~at J$^b$}
			      & \colhead{Norm.~by (7.7)$^c$}
			      & \colhead{$n$$^d$} 
			      & \colhead{Ellipticals$^e$} \\
\colhead{(1)}   & \colhead{(2)}  & \colhead{(3)}   & \colhead{(4)}  & \colhead{(5)}}
\startdata
 1.250\phn	&  1.000(0.000)\phn\phn  &  1.000(0.166)\phn\phn & 40\phn	&   1.000(0.000) \\ 
 1.650\phn	&  1.283(0.017)\phn\phn  &  1.364(0.201)\phn\phn & 40\phn	&   1.200(0.043) \\
 2.200\phn	&  1.093(0.026)\phn\phn  &  1.223(0.164)\phn\phn & 40\phn	&   0.940(0.038) \\
 2.469\phn	&  0.707(0.030)\phn\phn  &  0.794(0.111)\phn\phn & 40\phn	&   0.551(0.076) \\
 2.510\phn	&  0.683(0.037)\phn\phn  &  0.796(0.107)\phn\phn & 40\phn	&   0.493(0.050) \\
 2.550\phn	&  0.758(0.037)\phn\phn  &  0.843(0.123)\phn\phn & 40\phn	&   0.511(0.040) \\
 2.591\phn	&  0.773(0.033)\phn\phn  &  0.903(0.115)\phn\phn & 40\phn	&   0.547(0.002) \\
 2.631\phn	&  0.811(0.029)\phn\phn  &  0.889(0.128)\phn\phn & 40\phn	&   0.518(0.041) \\
 2.671\phn	&  0.823(0.037)\phn\phn  &  0.914(0.132)\phn\phn & 40\phn	&   0.532(0.032) \\
 2.712\phn	&  0.776(0.038)\phn\phn  &  0.926(0.114)\phn\phn & 40\phn	&   0.492(0.035) \\
\enddata                              
\tablenotetext{a}{The complete vesrion of this table is in the electronic edition of the Journal. 
		  The printed edition contains only a sample. The average spectra are shown in 
	          $f_{\nu}$, all normalized to have 1 Jy at J.  The standard deviation of the mean 
		  is given in the parentheses following each flux.}
\tablenotetext{b}{Individual spectra normalized at J.}
\tablenotetext{c}{Individual spectra normalized by AFE(7.7).}
\tablenotetext{d}{Number of galaxies used in the averaging.}
\tablenotetext{e}{The average spectrum of the 2 ellipticals using the J-band normalization scheme.
		  Note that the AFE(7.7) normalization scheme would lead to essentially the same 
		  average spectrum for the ellipticals.}
\end{deluxetable}

%
\begin{deluxetable}{rcll}
\tabletypesize{\scriptsize}
\tablenum{4}
\tablewidth{0pt}
\tablecaption{Infrared Lines and Emission Features}
\tablehead{
\colhead{Wavelength}  & & \colhead{Identifications\phn\phn\phn\phn} & \colhead{EW\phn\phn} \\
\colhead{($\mu$ms)}   & & \colhead{ }  &  \colhead{($\mu$m)\phn\phn}}
\startdata
             3.30\phn\phn\phn\phn &  \phn\phn\phn\phn   &       AFE(3.3)               &  0.02 \\
             4.03\phn\phn\phn\phn &  \phn\phn\phn\phn   &       Br$\alpha$?	      &  0.06 \\
             6.2\phn\phn\phn\phn &  \phn\phn\phn\phn   &        AFE(6.2)		      &  .... \\
             7.0\phn\phn\phn\phn &  \phn\phn\phn\phn   &        [Ar$\,$II]$\,6.99\mu$m $+$ AFE(6.9) $+$ H$_2$ 0-0 S(5) & 0.003 \\
             7.7\phn\phn\phn\phn &  \phn\phn\phn\phn   &        AFE(7.6/7.7)            &  .... \\
             8.6\phn\phn\phn\phn &  \phn\phn\phn\phn   &        AFE(8.6)   	      &  .... \\
            10.6\phn\phn\phn\phn &  \phn\phn\phn\phn   &        [S$\,$IV]$\,10.51\mu$m       &  0.002\\
            11.3\phn\phn\phn\phn &  \phn\phn\phn\phn   &        AFE(11.3)		      &  .... \\
\enddata
\end{deluxetable}

%
\begin{deluxetable}{lccrccc}
\tabletypesize{\scriptsize}
\tablenum{5}
\tablewidth{0pt}
\tablecaption{Median NIR-to-FIR Flux Ratios}
\tablehead{
\colhead{Subsample}  & \colhead{No.$^a$}
			      & \colhead{$R(60/100)^b$}
			      & \colhead{$\log\FtoB^b$}
			      & \colhead{NIR$^{\mathrm{total}}$/FIR$^b$}
			      & \colhead{NIR$^{\mathrm{ISM}}_{\mathrm{rc}}$/FIR$^b$}
			      & \colhead{NIR$^{\mathrm{ISM}}_{\mathrm{zr}}$/FIR$^b$} \\
\colhead{(1)}   & \colhead{(2)}  & \colhead{(3)}   & \colhead{(4)}  &
\colhead{(5)}   & \colhead{(6)}  & \colhead{(7)}}
\startdata
FIR-quiescent  &  12 &   0.39$\,$(0.06)  &  -0.18$\,$(0.24)\phn\phn & 0.174$\,$(0.10)   &   0.028$\,$(0.03)  &    0.041$\,$(0.03) \\
Intermediate   &  11 &   0.52$\,$(0.08)  &   0.37$\,$(0.35)\phn\phn & 0.083$\,$(0.06)   &   0.019$\,$(0.02)  &    0.035$\,$(0.02) \\
FIR-active     &  13 &   0.73$\,$(0.09)  &   0.54$\,$(0.39)\phn\phn & 0.043$\,$(0.04)   &   0.013$\,$(0.03)  &    0.020$\,$(0.03) \\
\enddata
\tablenotetext{a}{Actual number of galaxies in each subsample.}
\tablenotetext{b}{Median ratios, each followed in the parentheses by the r.m.s.~dispersion
		  with respect to the median. Columns~(5) to (7) are the ratios of the integrated 
                  PHT-S flux over 2.5 to 4.7$\,\mu$m to the FIR flux by treating the stellar continuum
		  differently: no stellar continuum is removed in Column~(5), and it is removed in both 
	          Columns~(6) and (7) using the reddening corrected and zero-reddening methods, respectively.}
\end{deluxetable}

%
\begin{deluxetable}{crccc}
\tabletypesize{\scriptsize}
\tablenum{6}
\tablewidth{0pt}
\tablecaption{Median Flux-Density Ratios of the AFEs}
\tablehead{
\colhead{No.$^a$}
			      & \colhead{$\log L_{FIR}/L_B^b$}
			      & \colhead{(6.2)/(7.7)$^b$} 
			      & \colhead{(8.6)/(7.7)$^b$} 
			      & \colhead{(11.3)/(7.7)$^b$} \\
\colhead{(1)}   & \colhead{(2)}  & \colhead{(3)}   & \colhead{(4)}  &
\colhead{(5)}}

\startdata
13  &-0.21(0.16)\phn\phn   & 0.69(0.12)  &    0.48(0.07) &  0.46(0.13) \\
13  & 0.20(0.15)\phn\phn   & 0.69(0.05)  &    0.48(0.08) &  0.47(0.11) \\
13  & 0.82(0.27)\phn\phn   & 0.65(0.04)  &    0.46(0.04) &  0.40(0.09) \\
All & 0.20(0.46)\phn\phn   & 0.66(0.09)  &    0.47(0.06) &  0.45(0.12) \\
\enddata
\tablenotetext{a}{Number of galaxies in each subsample.  The last row is for the whole sample.}
\tablenotetext{b}{Median ratios, each followed in parentheses by the r.m.s.~dispersion
		      with respect to that median.}
\end{deluxetable}

%
\begin{deluxetable}{lccc}
\tabletypesize{\scriptsize}
\tablenum{7}
\tablewidth{0pt}
\tablecaption{Median MIR-to-FIR Flux Ratios}
\tablehead{
\colhead{Subsample$^a$}       & \colhead{(5.8-11.3)/FIR$^b$}
			      & \colhead{AFEs(6-9)/FIR$^b$}
			      & \colhead{AFEs/FIR$^b$} \\
\colhead{(1)}   & \colhead{(2)}  & \colhead{(3)}   & \colhead{(4)}}
\startdata
FIR-quiescent   &  0.18$\,$(0.06)  & 0.13$\,$(0.05)  & 0.15$\,$(0.05) \\
Intermediate    &  0.17$\,$(0.04)  & 0.13$\,$(0.04)  & 0.15$\,$(0.04) \\
FIR-active      &  0.11$\,$(0.05)  & 0.09$\,$(0.04)  & 0.09$\,$(0.04) \\
\enddata
\tablenotetext{a}{Same subsamples as in Table~5.} 
\tablenotetext{b}{Column~(2) is the median ratio of the integrated flux
		      between 5.8 and 11.3$\,\mu$m to the FIR flux.  Column~(3) is the median ratio of the summed flux 
		      of AFE(6.2), AFE(7.7) and AFE(8.6) to the FIR flux.  Column~(4) is the median ratio of 
		      the summed flux of all the 4 MIR features to the FIR flux.  Each median flux ratio is followed 
		      in the parentheses by the r.m.s.~dispersion with respect to that median.  PHT-S aperture correction
	              has been applied to all the MIR fluxes.}
\end{deluxetable}

%
\begin{deluxetable}{lccc}
\tabletypesize{\scriptsize}
\tablenum{C1}
\tablewidth{0pt}
\tablecaption{Feature Profile Slopes}
\tablehead{
\colhead{Profile Slope$^a$}  & \colhead{Median} & \colhead{Dispersion$^b$}  & \colhead{$\sigma_m^c$} \\
\colhead{(1)}   & \colhead{(2)}  & \colhead{(3)}   & \colhead{(4)}}
\startdata
\phn\phn$S_{-}(6.2)$    &  0.69  & 0.14  & 0.05  \\
\phn\phn$S_{+}(6.2)$    &  0.42  & 0.14  & 0.06  \\
\phn\phn$S_{-}(7.7)$    &  0.49  & 0.08  & 0.03  \\
\phn\phn$S_{+}(7.7)$    &  0.33  & 0.08  & 0.03  \\
\phn\phn$S_{-}(8.6)$    &  0.09  & 0.12  & 0.04  \\
\phn\phn$S_{+}(8.6)$    &  0.55  & 0.15  & 0.10  \\
\phn\phn$S_{-}(11.3)$   &  0.58  & 0.24  & 0.11  \\
\enddata
\tablenotetext{a}{$S_-$ and $S_+$ refer to the profile slopes at the short- and long-wavelength sides of an 
          emission feature, respectively.}
\tablenotetext{b}{The r.m.s~dispersion with respect to the median.}
\tablenotetext{c}{Median of the statistical measurement errors in the profile slope.}
\end{deluxetable}


\begin{figure}
\plotone{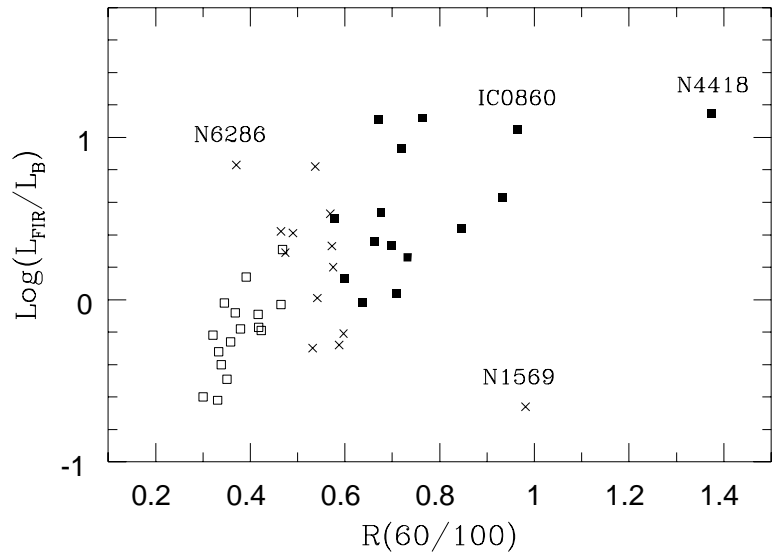}
\caption{
     Distribution of the galaxies in a plot of $\log\FtoB$
     {\it vs.} $R(60/100)$.  The sample is
     divided into an ``FIR-quiescent'' subsample (open squares), an ``FIR-active''
     subsample (solid squares), and an intermediate subsample (crosses). 
     NGC 3620, for which $L_B$ is not available,
     was assigned to the intermediate subsample, and is omitted from 
     Figs.~1, 7, and 8. 
     }
\end{figure}
\newpage

\begin{figure}
\plotone{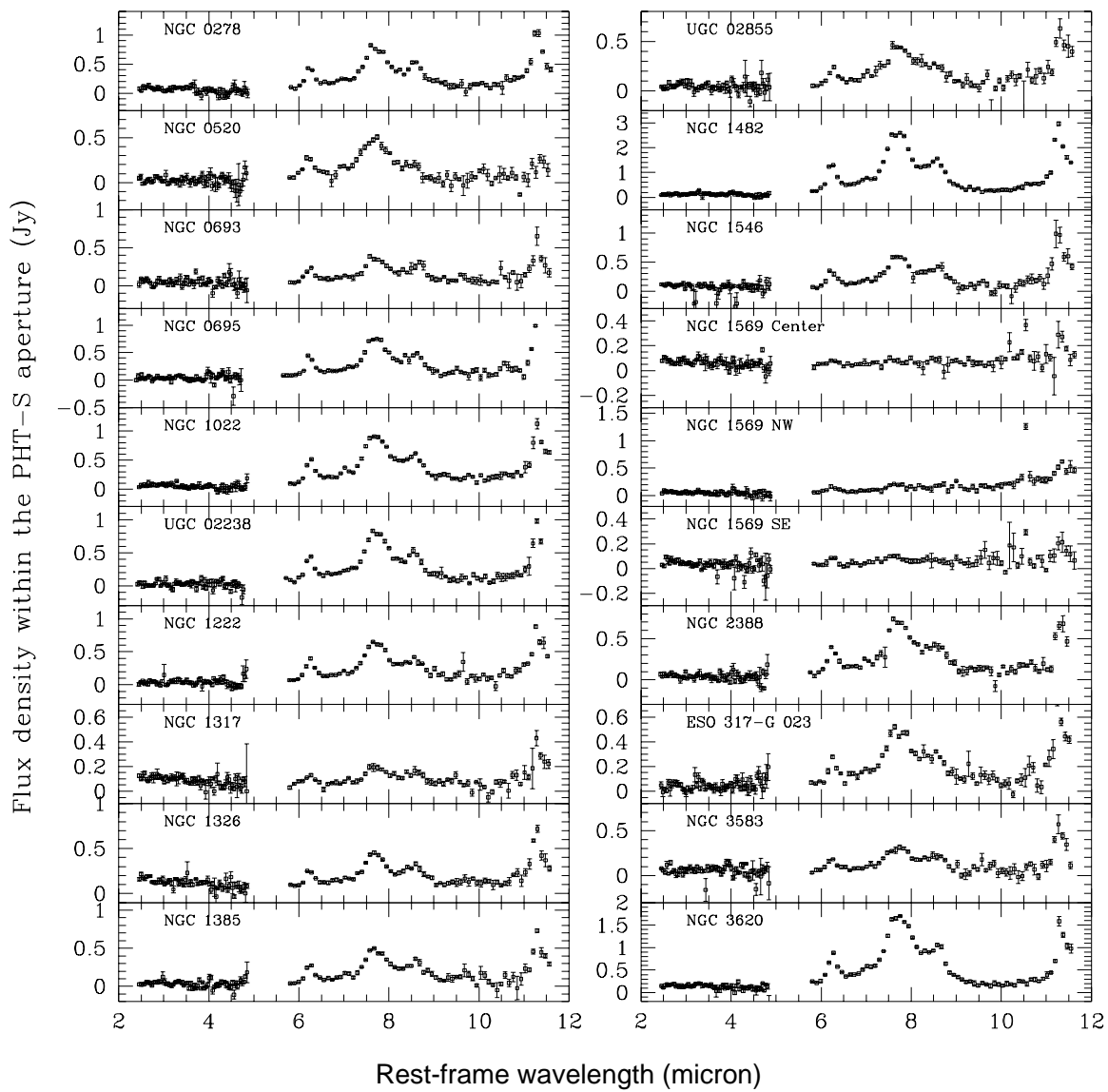}
\caption{
     PHT-S spectra in units of Jy as a function of the rest-frame wavelength
     in microns. The error bars show the statistical errors for each spectral point.
     }
\end{figure}
\newpage

\setcounter{figure}{1}
\begin{figure}
\plotone{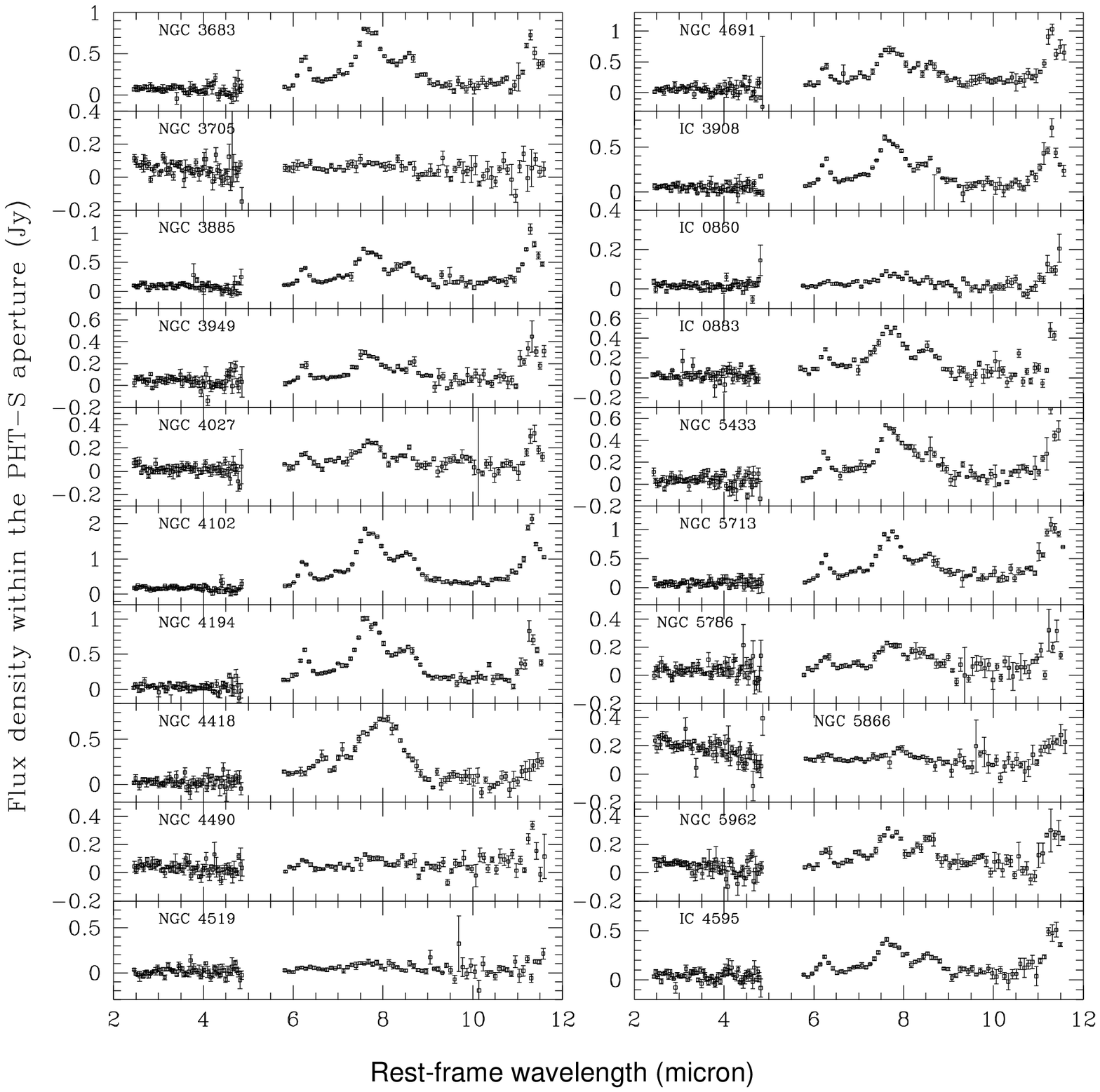}
\caption{
     Continuation of Fig.~2$a$.
     }
\end{figure}
\newpage

\setcounter{figure}{1}
\begin{figure}
\plotone{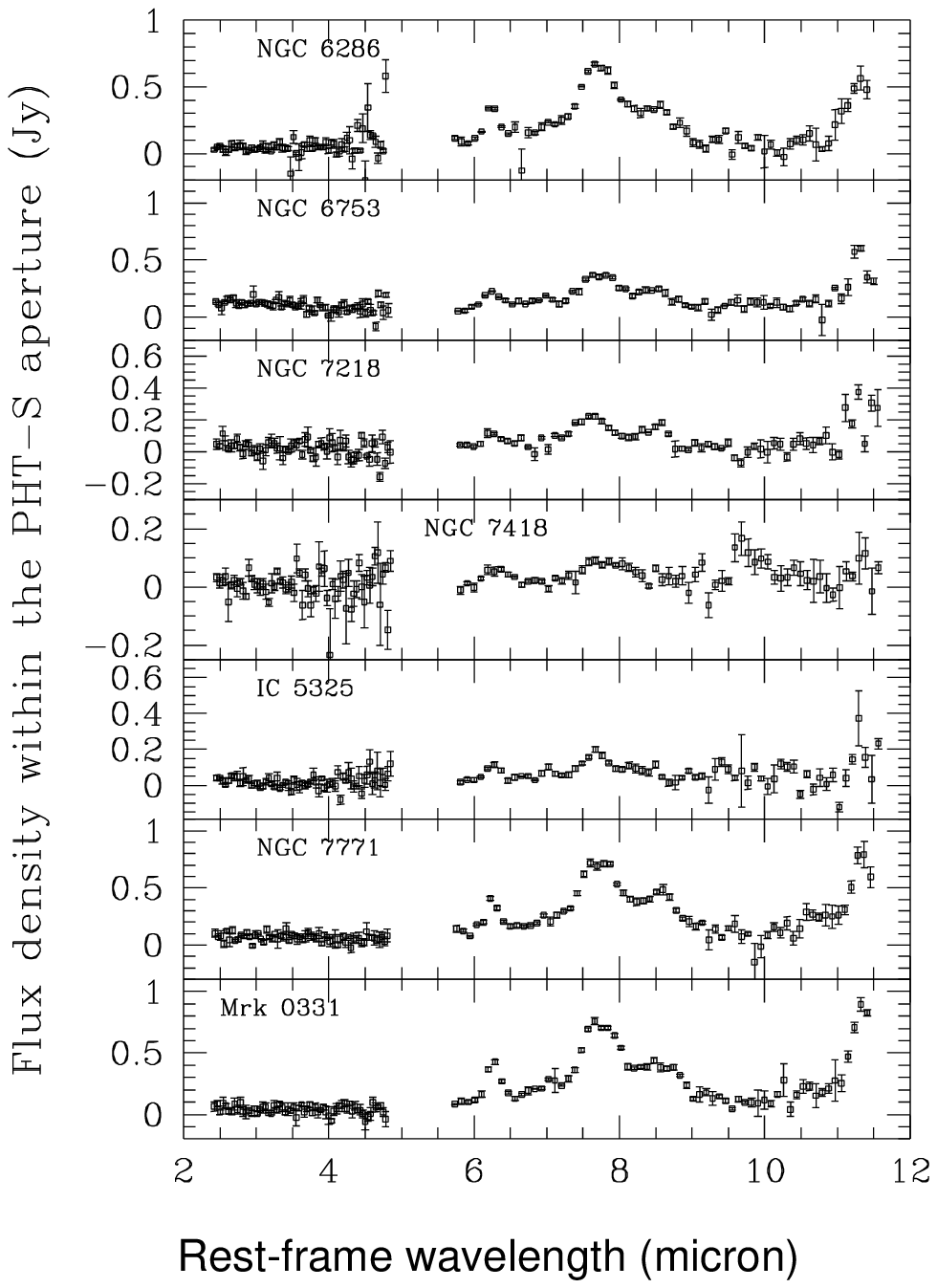}
\caption{
     Continuation of Fig.~2$a$.
     }
\end{figure}
\newpage

\begin{figure}
\plotone{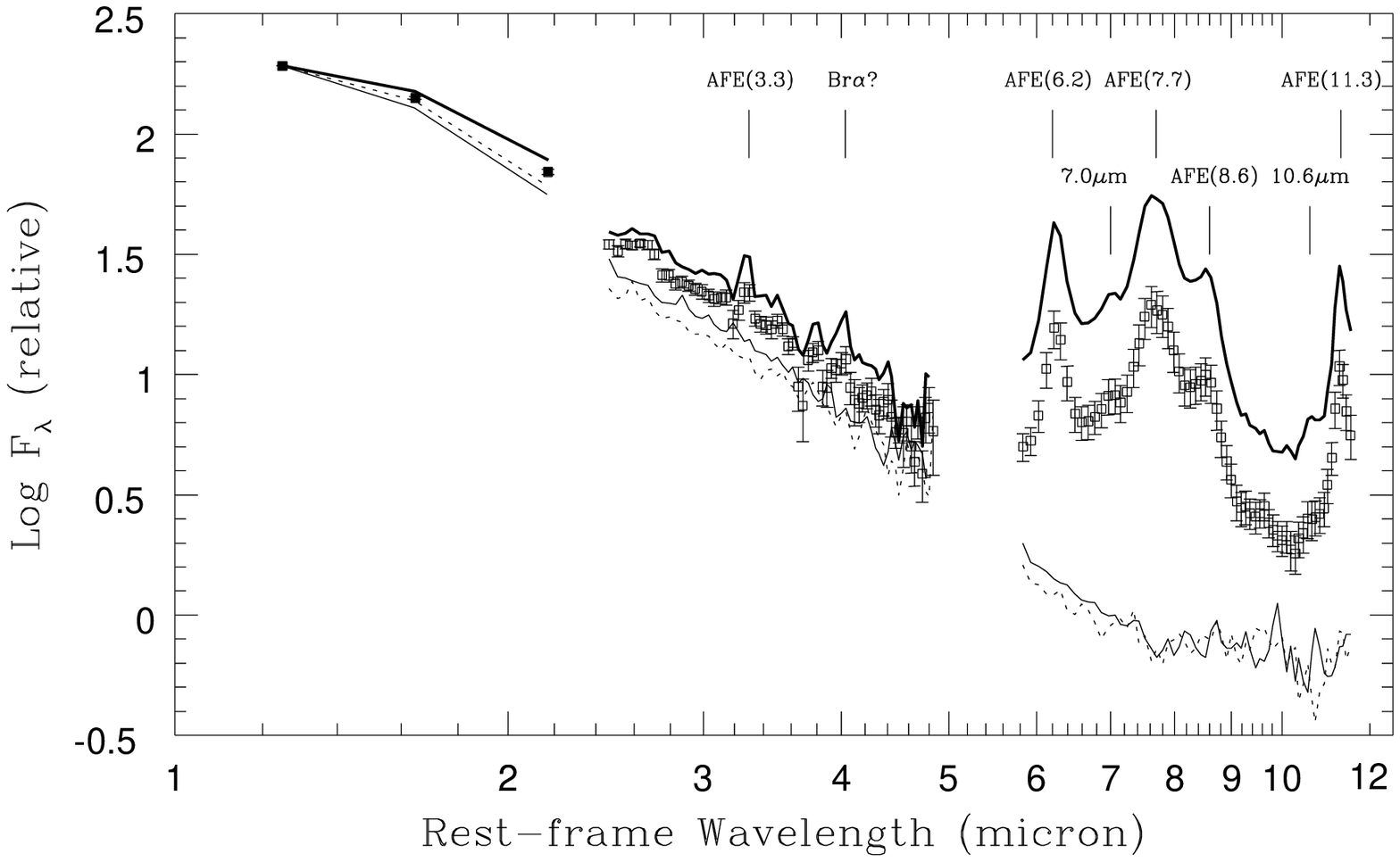}
\vspace{-1.7in}
\caption{
     Plots of the average rest-frame spectra derived 
     from weighted averages of the spectra 
     of 40 galaxies (see \S 3.1).
     The squares represent the average spectrum obtained by normalizing
     to the J fluxes, while the thick solid curve results from normalizing
     by the integrated flux of AFE(7.7\um). Representative error bars
     are shown only for the former; note that the error bars should
     smallest near the fiducial wavelength used for the normalization.
     The spectra of two elliptical galaxies are also shown: 
     NGC$\,$3379 (the thin solid curve) and NGC$\,$4374
     (the dotted line). 
     }
\end{figure}
\newpage

\begin{figure}
\plotone{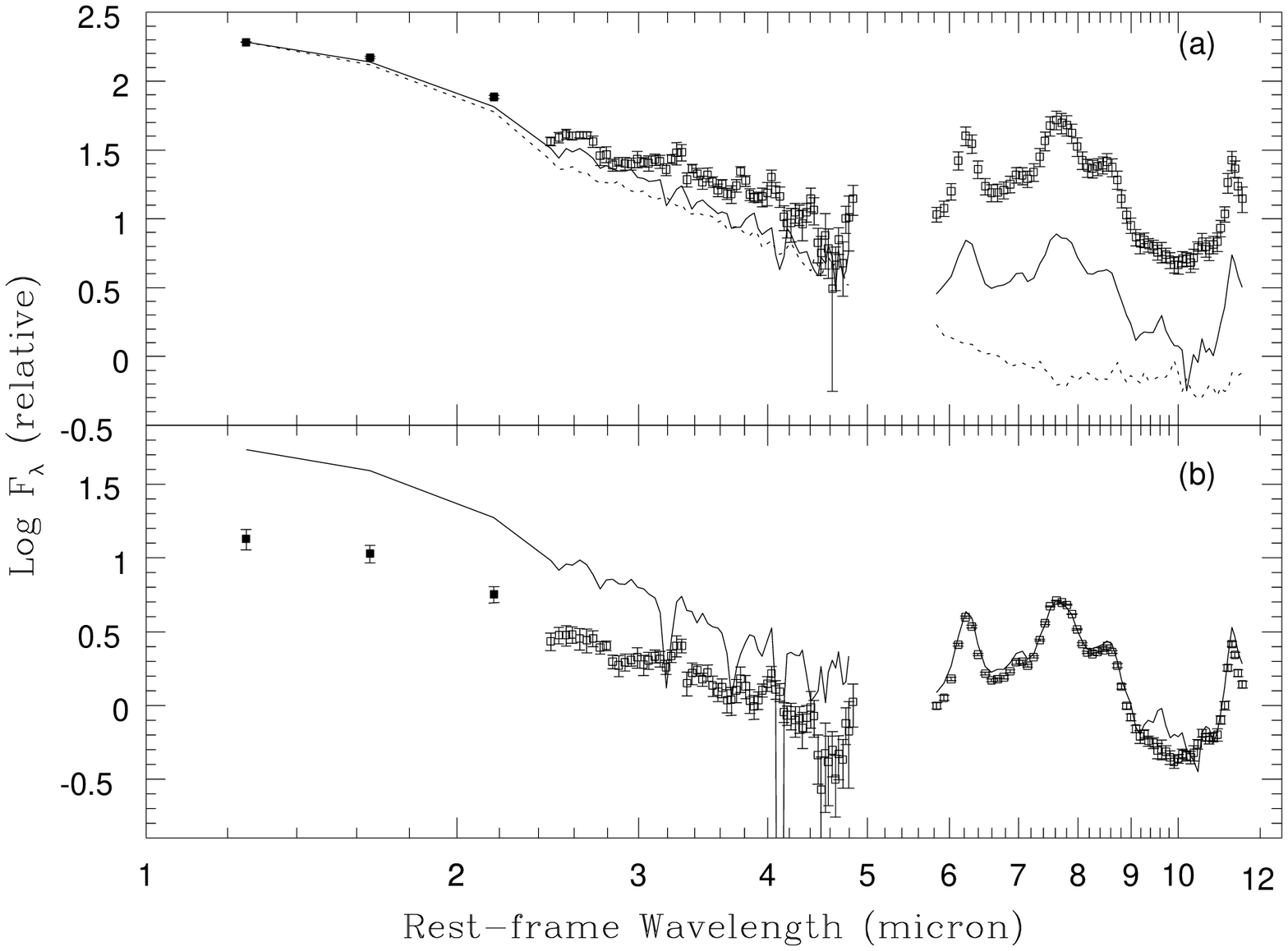}
\vspace{-0.5in}
\caption{
     Comparison of the average spectra of the FIR-quiescent and FIR-active 
     subsamples (see \S 2.1; also Table~5). The squares represent 
     the FIR-active subsample and the solid curve is the average spectrum 
     of the FIR-quiescent galaxies.  The mean of the elliptical galaxy 
     spectra is shown as a dotted curve. (Numerical values for all three 
     curves can be found in Table~3.)  As explained in the text, different 
     normalizations are used for generating the spectra in (a) and (b).
     }
\end{figure}
\newpage

\begin{figure}
\plotone{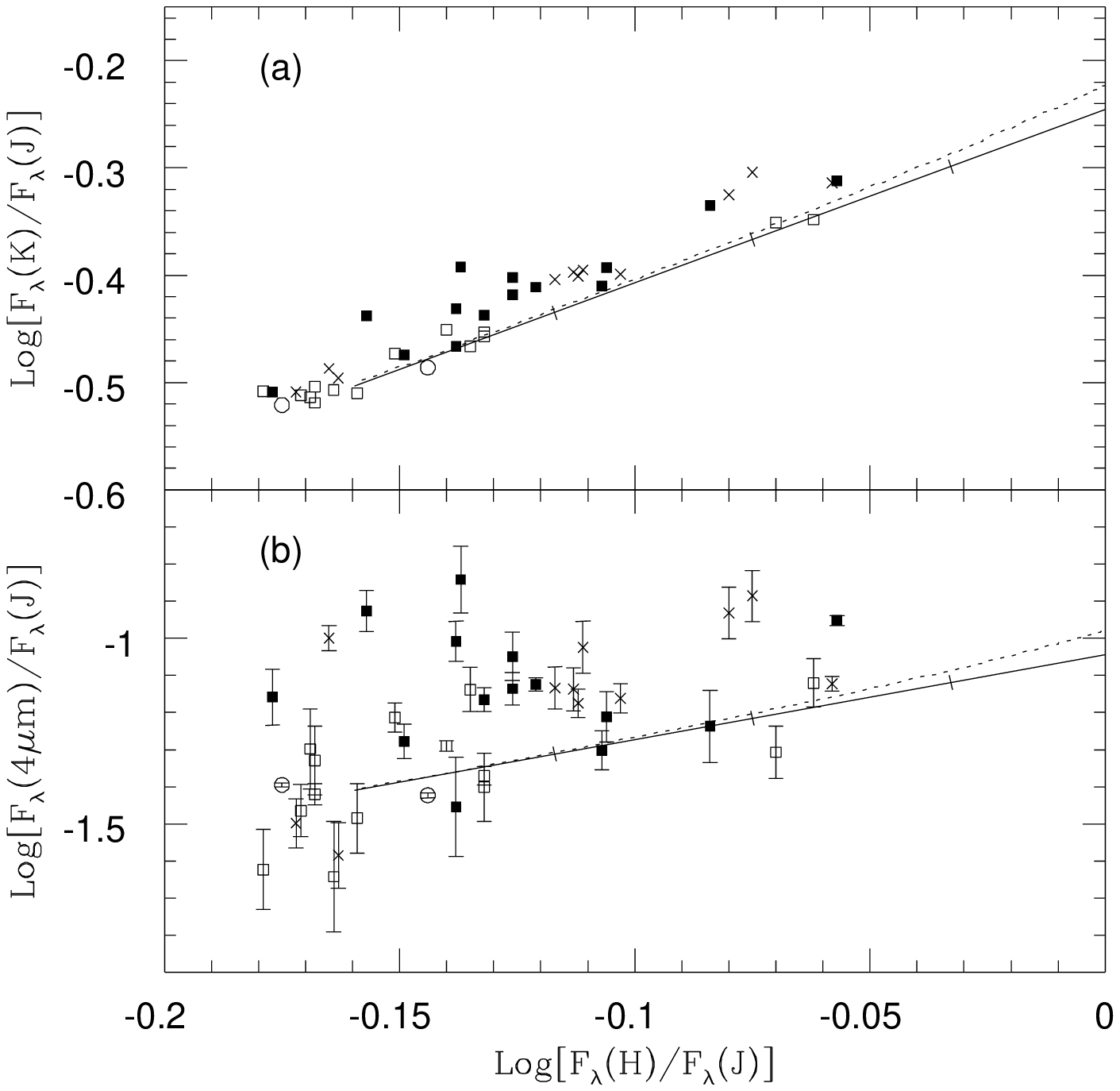}
\caption{
       Near-IR ``color-color'' plots: $(a)$ log $F_{\lambda}(K)/F_{\lambda}(J)$ 
       {\it vs.} 
       log $F_{\lambda}(H)/F_{\lambda}(J)$, and 
       $(b)$ log $F_{\lambda}(4\mu{\rm m})/F_{\lambda}(J)$ 
       {\it vs.} log $F_{\lambda}(H)/F_{\lambda}(J)$, 
       where $F_{\lambda}(4\mu{\rm m})$ is the mean flux density at 
       4$\mu$m as defined in Table~2.  The Key Project galaxies are shown 
       as open squares, solid squares, and crosses (as in Fig.~1), 
       and the reference elliptical galaxies as circles.
       In each panel, the solid line is the inferred 
       reddening line for the case of a foreground dust screen 
       the dotted curve is for the case where dust and stars
       are uniformly mixed.  The tick marks along the solid line indicate 
       respectively A$_{V}$ = 1, 2, and 3$^{m}$ for the dust screen case.
	}
\end{figure}
\newpage

\begin{figure}
\plotone{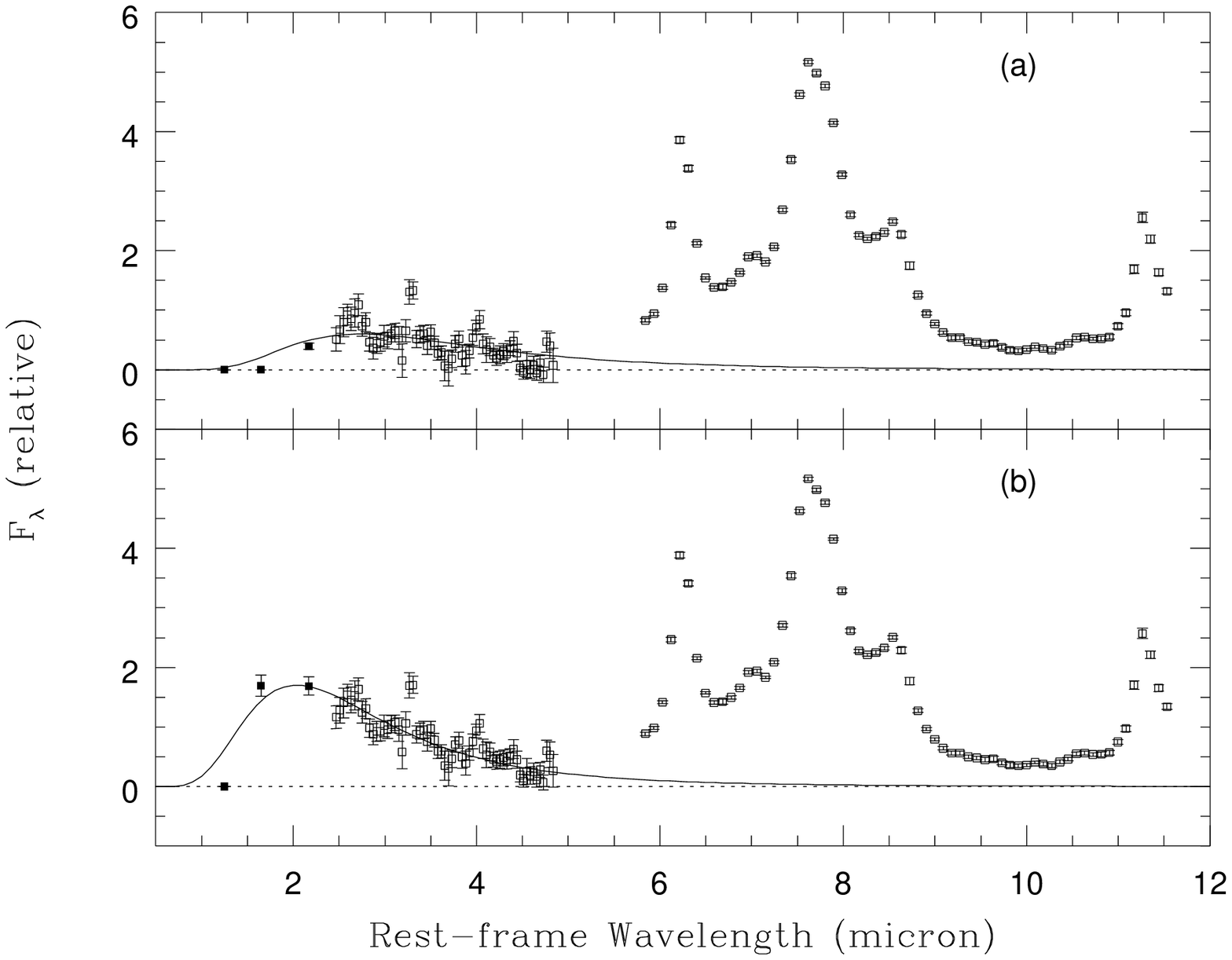}
\vspace{-1.0in}
\caption{
       Averaged spectra obtaining by subtracting from each disk galaxy spectrum
       $(a)$ a stellar continuum reddened to agree with the H/J color, or
       $(b)$ an unreddened stellar continuum (derived from the elliptical galaxies.)
       The solid curves are  
       modified black-bodies with an emissivity that scales as 
       $\lambda^{-2}$ and $T = 750\,$K in $(a)$ and $10^3\,$K in $(b)$. The error bars 
       represent the standard deviation of the mean.  The individual, stellar
       continuum-subtracted spectra were normalized by the integrated flux of 
       AFE(7.7) prior to averaging, and the results are renormalized to the same peak
       flux of the 7.7\um\ feature.
     }
\end{figure}
\newpage

\begin{figure}
\plotone{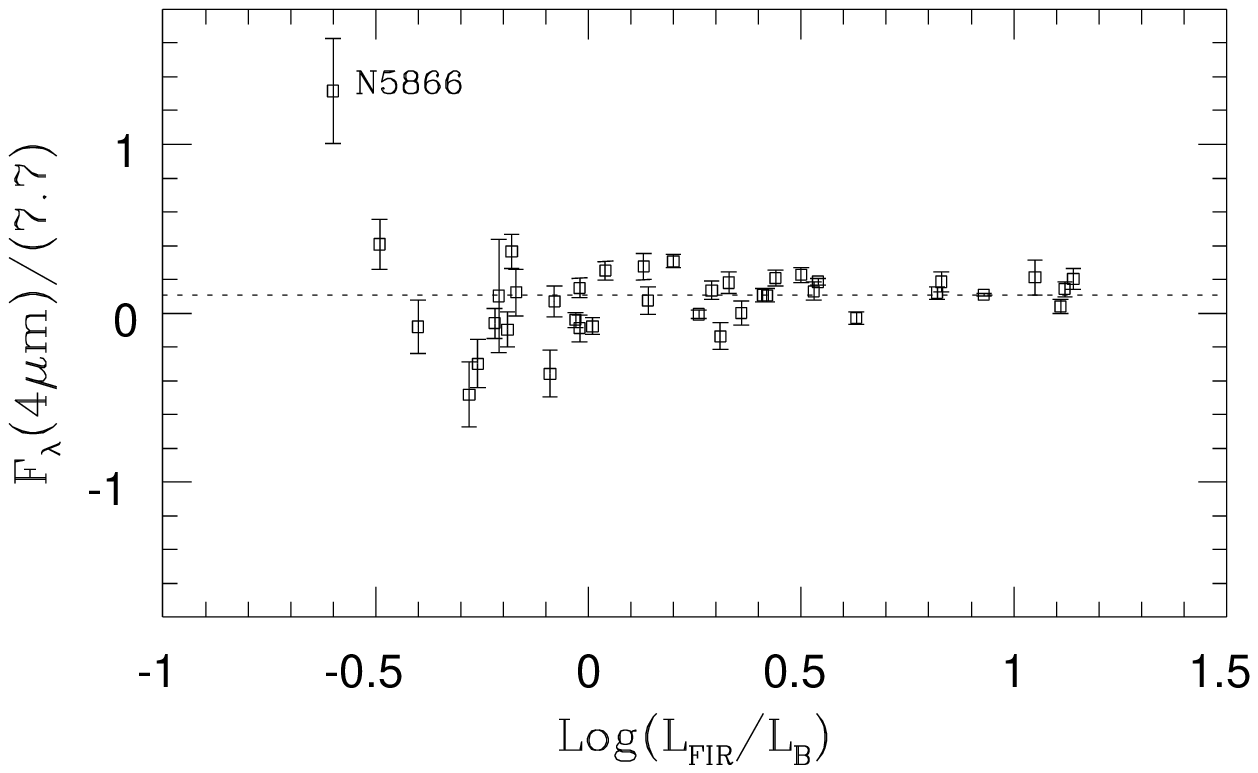}
\vspace{-1.0in}
\caption{
        Plot of the ratio of 
        the flux density at 4$\mu$m to the mean flux density AFE(7.7),  
        as a function of \FtoB.
        Both $F_{\lambda}$(4\um) and (7.7) were computed
        after removal of the stellar contributions.
        The dotted line indicates the median sample flux ratio of 0.11. 
     }
\end{figure}
\newpage

\clearpage

\begin{figure}
\plotone{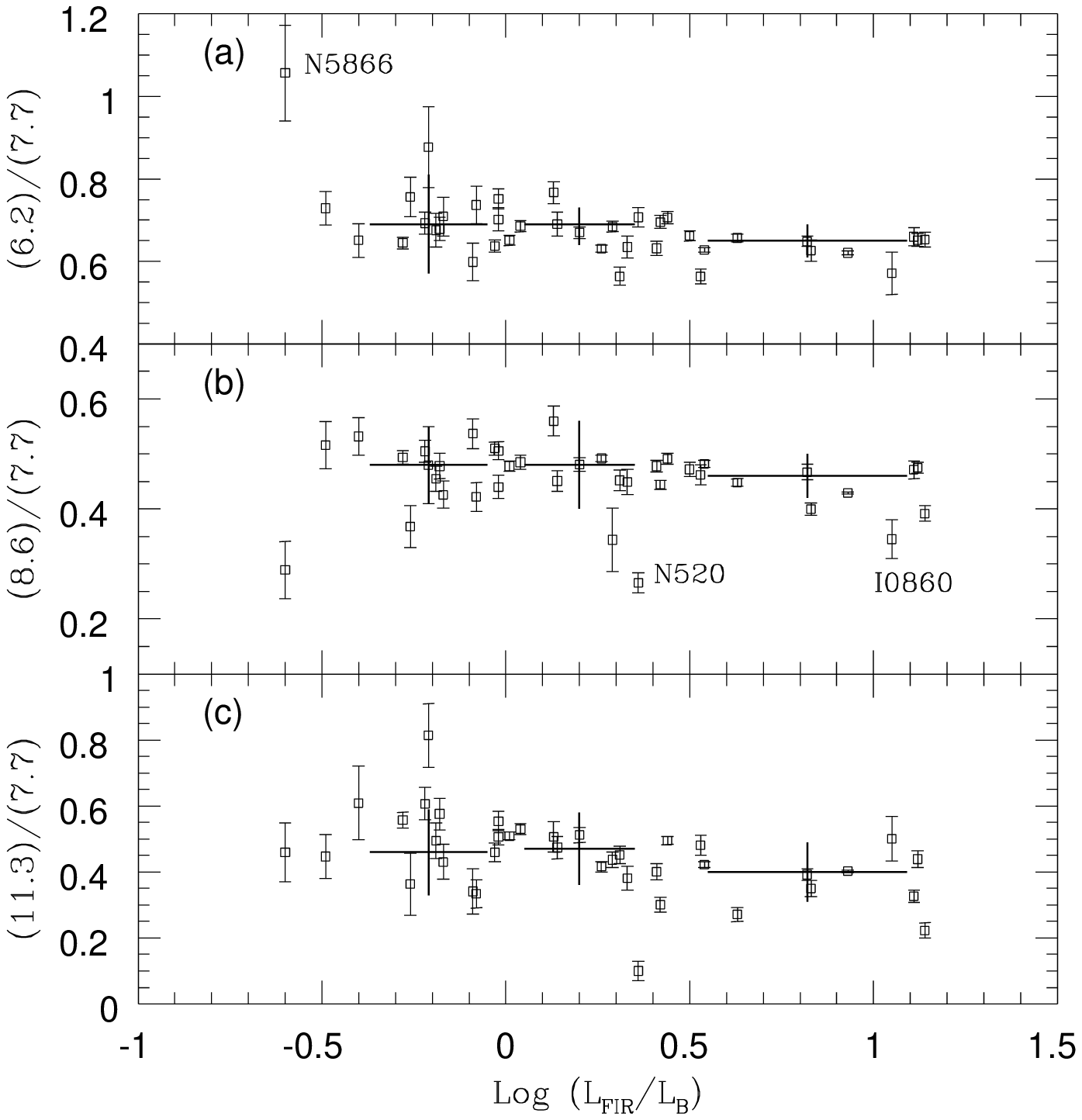}
\caption{
     Plots of the relative strengths of the AFEs as a function
     of \FtoB, where (6.2)/(7.7), (8.6)/(7.7), (11.3)/(7.7) 
     represent the mean $F_{\lambda}$ for these features after removal
     of the stellar continuum (using the zero-reddening method).
     We have divided the data
     set into three equal-size subsamples 
     in terms of \FtoB, and plotted their 
     median values as large crosses in each plot.  The extent of each cross indicate 
     the r.m.s.~dispersion with respect to the median value for each 
     subsample.  
     A few outliers are labelled.
     }
\end{figure}
\newpage

\begin{figure}
\plotone{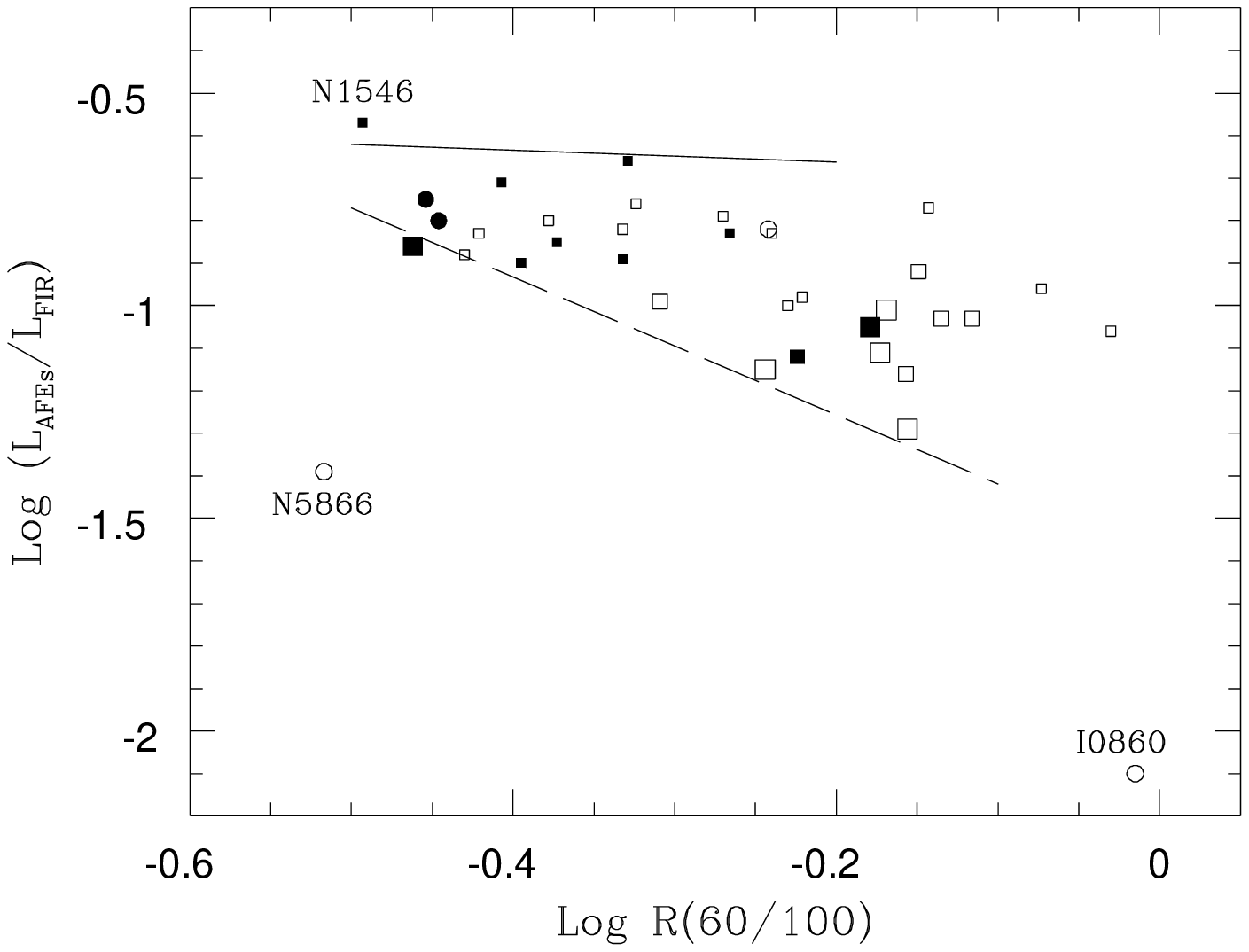}
\vspace{-1.0in}
\caption{
     Plot of the ratio of the combined luminosity of the AFEs to the FIR
     dust emission, as a function of $R(60/100)$.  
     Galaxies with (without) a measured interstellar far-UV radiation field $G_0$ in 
     Malhotra \etal (2001) are represented by squares (circles). 
     The sizes of the squares represent the approximate value of $G_0$, 
     and the filled symbols represent galaxies with the largest aperture corrections,
     as described in the text. 
     The statistical errors in \LAFE/\LFIR\ 
     are small and thus no error bars are plotted.
     Note that nearly all of the galaxies
     fall within a wedge defined by the solid and dotted lines, derived from {\it IRAS}
     observations of reflection nebulosity and \ion{H}{2} regions respectively.
     }
\end{figure}
\newpage

\begin{figure}
\plotone{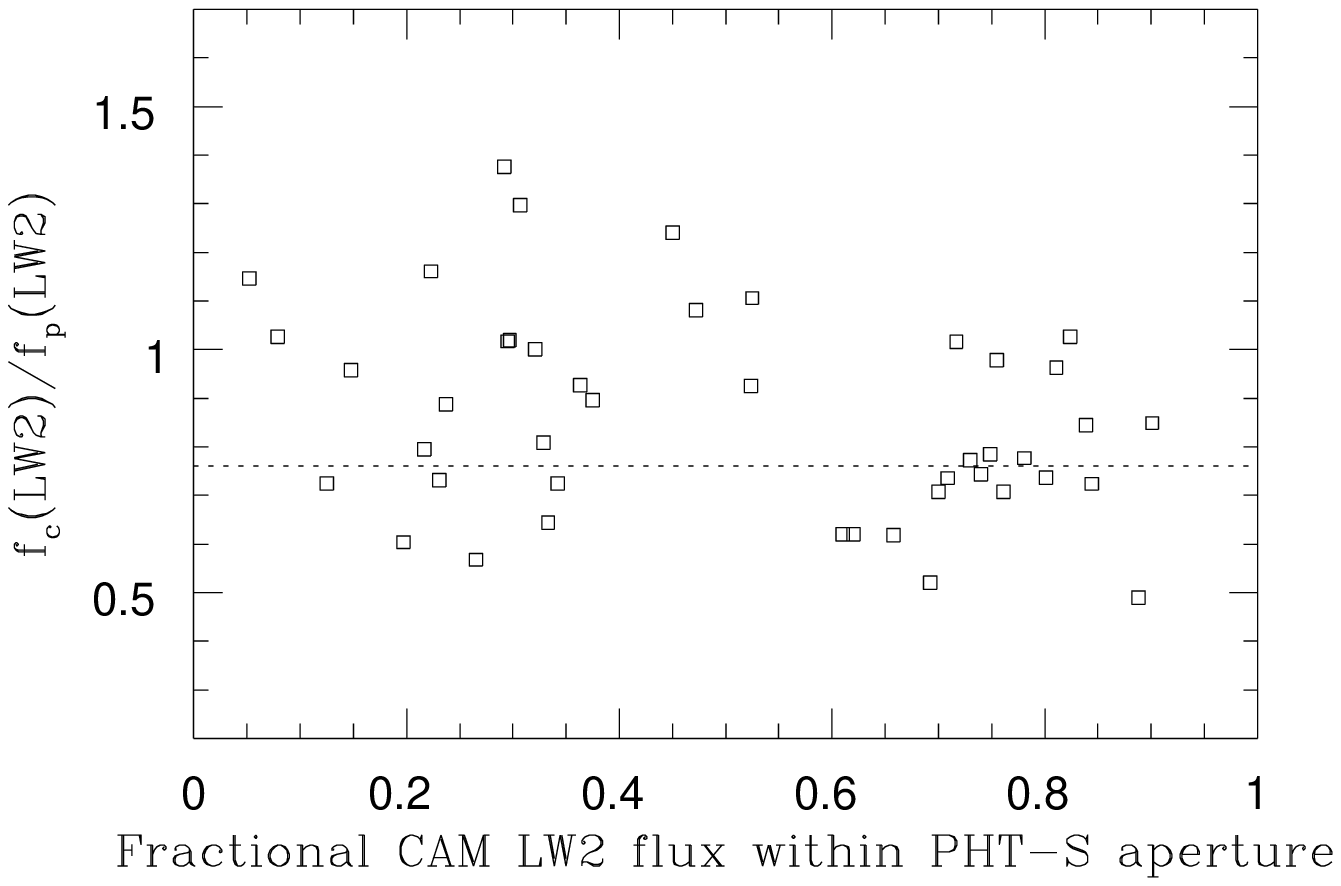}
\caption{
     Comparison of the PHT-S flux scale with the flux scale of CAM in the 
     LW2 filter ($\lambda_0 = 6.7\um$).  The abscissa is $p$, the fraction of 
     the galaxy flux at 6.7\um\ that falls within the PHT-S $24''\times 24''$
     aperture. The ordinate is the ratio of 
     the CAM LW2 flux within the PHT-S aperture to a 
     CAM LW2-band-equivalent flux derived from the PHT-S spectrum.  
     The dotted line
     indicates a ratio of 0.76, the mean value for the 20 
     sources with $p > 0.6$.
     }
\end{figure}
\newpage

\begin{figure}
\plotone{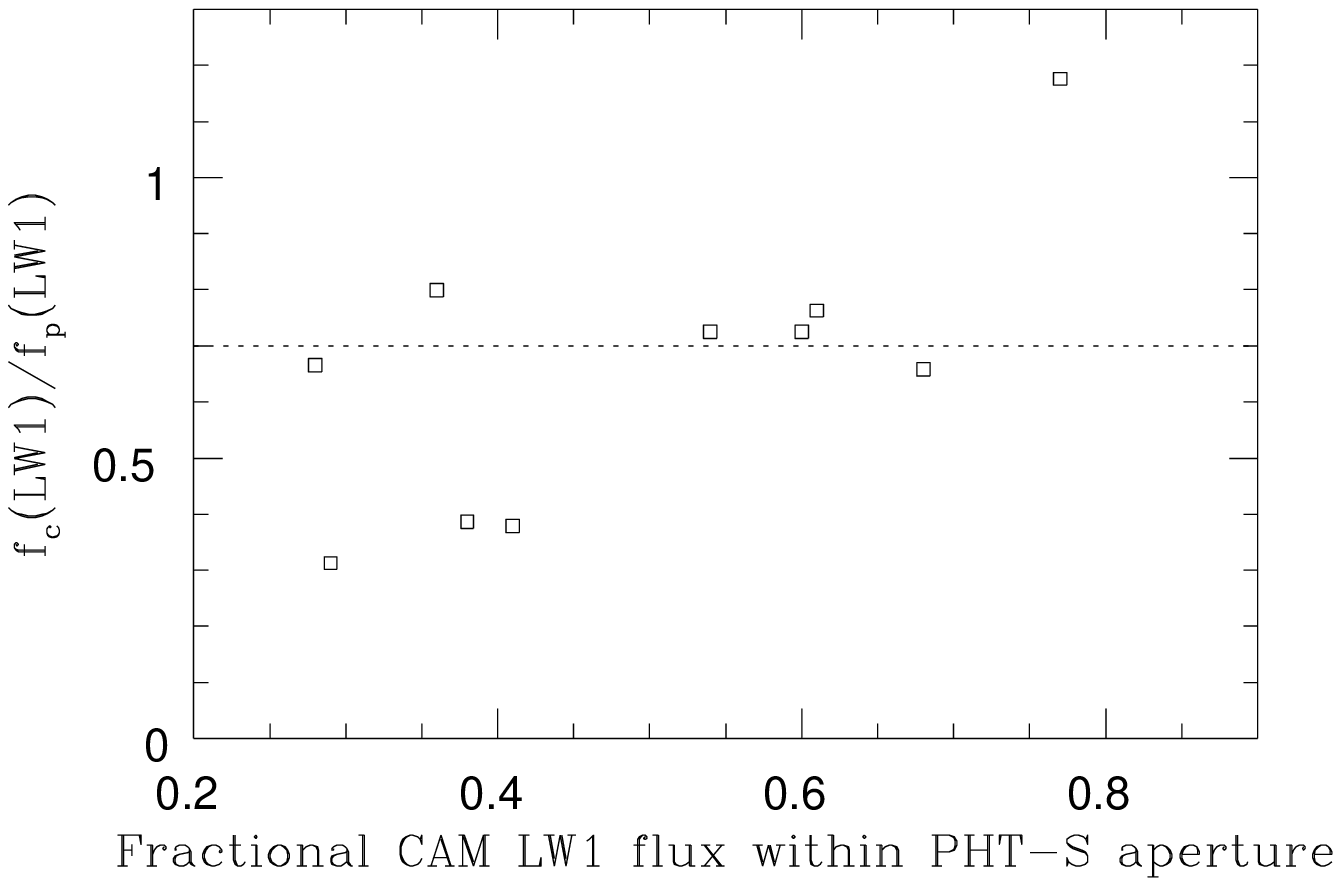}
\caption{
     Similar to Fig.~10, but comparing with CAM LW1 (4 to 5\um) data. The dotted line
     indicates a flux ratio of 0.7, the median value of the data shown in the plot.
     }
\end{figure}
\newpage

\begin{figure}
\plotone{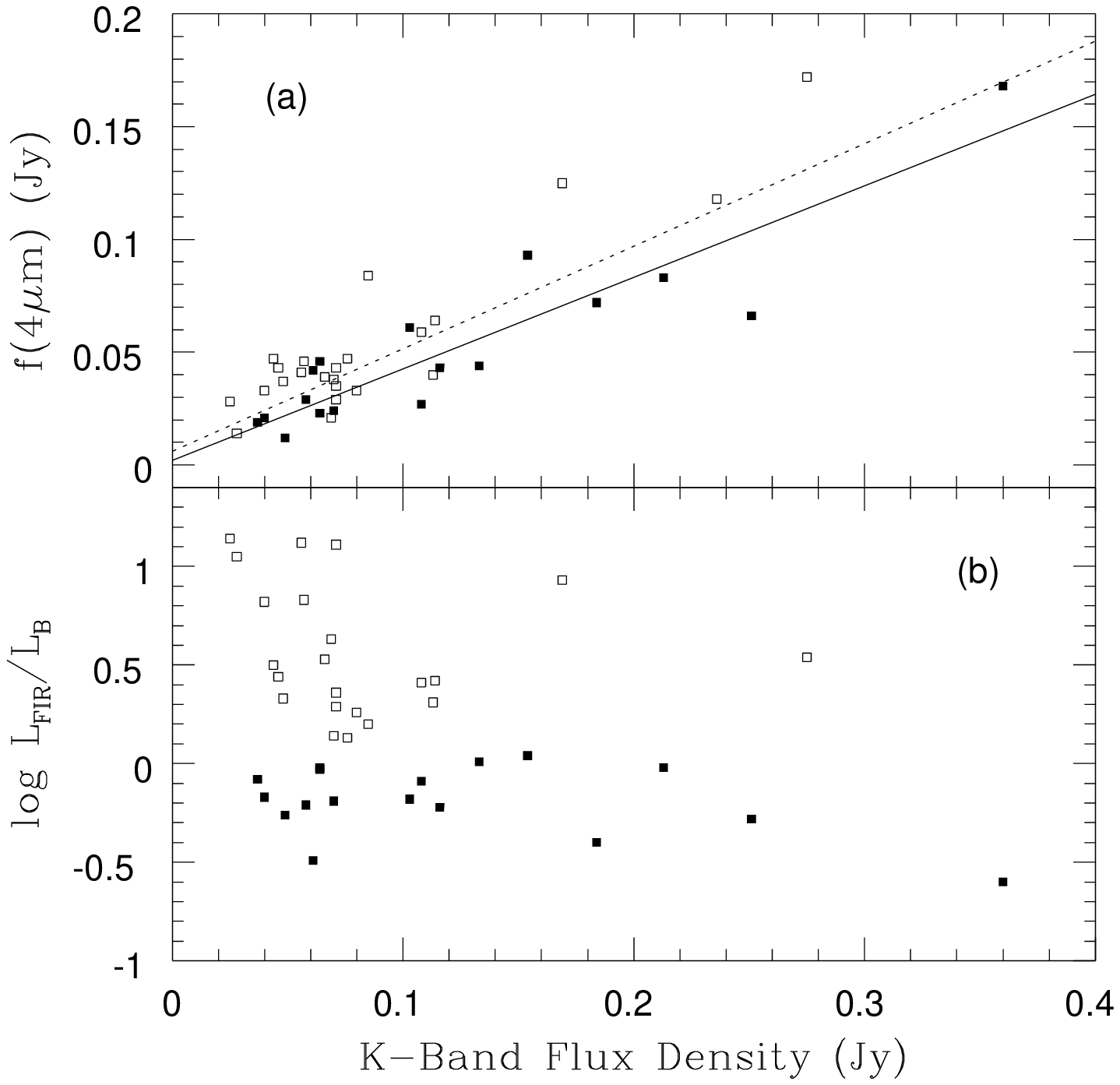}
\vspace{-1.5in}
\caption{
     Plots of $(a)$ the flux density at 4$\mu$m and $(b)$ the FIR-to-blue
     luminosity ratio as a function of K-band flux density for 40 galaxies.
     The galaxies with $\log\FtoB < 0.1$ are represented by solid squares;  
     notice that these galaxies have a flatter distribution in both plots,
     especially in $(b)$, where they show no trend with increasing F$_K$.
     The dotted line in $(a)$ is a least-squares fit to all 40 galaxies, by minimizing
     their distances perpendicular to the fit. The solid line is a similar fit
     to the galaxies indicated by solid squares.
     }
\end{figure}

\end{document}